\documentclass[useAMS,usenatbib,usegraphicx,mathabx]{mn2e}

\usepackage{amsmath}
\DeclareRobustCommand{\ion}[2]{%
\relax\ifmmode
\ifx\testbx\f
{\mathrm{#1\,\textsc{#2}}}\else
{\mathrm{#1\,\mathsc{#2}}}\fi
\else\textup{#1\,{\mdseries\textsc{#2}}}%
\fi}

\title[AO imaging and optical spectroscopy of the Bird] 
      {Adaptive optics imaging and optical spectroscopy 
      of a multiple merger in a luminous infrared galaxy$^{1}$
     }

\author[P. V\"ais\"anen et al.]
       {P. V\"ais\"anen,$^{1}$
          S. Mattila,$^{2,3}$ 
          A. Kniazev,$^{1}$
          A. Adamo,$^{4}$ 
          A. Efstathiou,$^{5}$ 
          D. Farrah,$^{6}$ \newauthor
          P.H. Johansson,$^{7}$
          G. \"Ostlin,$^{4}$ 
          D.A.H. Buckley,$^{1}$
          E.B. Burgh,$^{8}$
          L. Crause,$^{1}$\newauthor
          Y. Hashimoto,$^{1}$
          P. Lira,$^{9}$
          N. Loaring,$^{1}$
          K. Nordsieck,$^{8}$        
          E. Romero-Colmenero,$^{1}$\newauthor
          S. Ryder,$^{10}$
          M. Still,$^{1}$
          A. Zijlstra$^{11,1}$
          \\ 
          $^{1}$South African Astronomical Observatory, P.O.Box 9,
                Observatory, 7935, Cape Town, South Africa (petri@saao.ac.za)\\
          $^{2}$Tuorla Observatory, University of Turku, V\"ais\"al\"antie 20,
                FI-21500 Piikki\"o, Finland \\
          $^{3}$Astrophysics Research Centre, School of Mathematics and 
                Physics, Queen's University of Belfast, Belfast BT7 1NN\\
          $^{4}$Department of Astronomy, Stockholm University, AlbaNova, 
                SE-106 91 Stockholm, Sweden\\
          $^{5}$School of Computer Science and Engineering, Cyprus College, 
                Engomi, 1516 Nicosia, Cyprus\\
          $^{6}$Department of Astronomy, Cornell University, Ithaca, NY 14853, 
                USA\\
          $^{7}$Universit\"ats-Sternwarte M\"unchen, Scheinerstr. 1., D-81679
                M\"unchen, Germany\\
          $^{8}$Space Astronomy Laboratory, University of Wisconsin, Madison,
                WI 57306, USA\\
          $^{9}$Departamento de Astronomía, Universidad de Chile, Santiago, 
                Chile\\
          $^{10}$Anglo-Australian Observatory, Epping, NSW 1710, Australia\\
          $^{11}$Jodrell Bank Center for Astrophysics, 
                 School of Physics and Astronomy, University of Manchester,  
                 Oxford St, Manchester M13 9PL}

\date{Accepted ... Received ..; in original form ..}

\pagerange{\pageref{firstpage}--\pageref{lastpage}} \pubyear{2007}

\begin{document}

\maketitle

\label{firstpage}

\begin{abstract}

We present near-infrared (NIR) adaptive optics imaging obtained with 
VLT/NACO and optical spectroscopy from the Southern African
Large Telescope (SALT) of a luminous infrared galaxy (LIRG) 
IRAS~19115-2124.  These data are combined with archival {\em HST} imaging
and {\em Spitzer} imaging and spectroscopy, allowing us to 
study this disturbed interacting/merging galaxy, dubbed the Bird, 
in extraordinary detail.  In particular, the data reveal a 
triple system where the LIRG phenomenon is dominated by the smallest
of the components.

One nucleus is a regular barred spiral with significant 
rotation, 
while another is highly disturbed with a surface brightness 
distribution intermediate to that of disk and bulge systems, 
and hints of remaining arm/bar structure.  
We derive dynamical masses in the range 3--7$\times 10^{10} {\rm M}_{\odot}$
for both. The third component appears to be a 
1--2$\times 10^{10} {\rm M}_{\odot}$ mass irregular galaxy.
The total system exhibits \ion{H}{II} galaxy-like optical line 
ratios and strengths, and no evidence for AGN activity is found from 
optical or mid-infrared data.  
The star formation rate is estimated to be $\sim 190\ {\rm M_{\odot} yr^{-1}}$.
We also report a search for supernovae from NIR images separated by 5 months
and search for super star cluster candidates.
We detect outflowing gas from the Bird mostly in the range 
100--300 km~s$^{-1}$ using \ion{Na}{I} D absorption features.
Overall, the Bird shows kinematic, dynamical, and 
emission line properties typical for cool ultra luminous IR galaxies. 
However, the interesting features setting it apart for future studies 
are its triple merger nature, and the location of its star formation peak -- 
the strongest star formation, as revealed by {\rm Spitzer} imaging, 
does not come from the two major $K$-band nuclei, but from the 
third irregular component.  This is in 
contrast to the conventional view that the (U)LIRG phases are powered by 
infalling gas to the major nuclei 
of the merging spiral galaxies. 
Aided by simulations, we discuss scenarios where the irregular component 
is on its first high-speed encounter with the more massive components. 

\end{abstract}


\section{Introduction}

\footnotetext[1]
{
 Based on observations made with ESO Telescopes at the 
      Paranal Observatory under programme 073.D-0406A, and with
      the Southern African Large Telescope (SALT).}

Over the last three decades, evidence has mounted that luminous infrared 
(IR) galaxies (LIRGs\footnote[2]{objects with 8--1000$\mu$m luminosities in 
excess of $10^{11}{\rm L}_{\odot}$; objects with 8--1000$\mu$m
luminosities $>10^{12}{\rm L}_{\odot}$ 
are commonly referred to as ultraluminous IR galaxies, ULIRGs}) 
may signpost important events in understanding the wider 
picture of galaxy formation and evolution. First hinted at by ground based 
studies in the 1970s \citep{rie}, the importance of infrared-luminous 
activity in galaxies was clearly demonstrated by surveys with {\em IRAS} 
\citep{hou85}, which found large 
numbers of extragalactic sources that were bright at mid/far-IR wavelengths, 
and varied substantially in nature as a function of infrared (8--1000$\mu$m) 
luminosity. Those sources with IR luminosities less than about 
$10^{10}{\rm L}_{\odot}$ 
were mainly either dwarf galaxies or relatively dustless 
ellipticals, where the IR emission arose almost entirely from dust grains 
heated by ambient interstellar light. Sources with IR luminosities in the 
range $10^{10}{\rm L}_{\odot}<L_{ir}<10^{11.5}{\rm L}_{\odot}$ 
contained a significant 
fraction of ellipticals, but with a rising number of IR-luminous disk 
galaxies and galaxy mergers. At IR luminosities of $>10^{11.5}{\rm L}_{\odot}$ 
however, mergers were almost ubiquitous, with very few undisturbed disk or 
elliptical systems. In the bulk of these systems the IR emission was thought 
to arise from some combination of dense, compact starbursts, and deeply 
buried AGN. Surveys with {\em ISO} \citep[e.g.][]{elbaz} and 
SCUBA \citep[e.g.][]{hug,cop} 
then showed that LIRGs become increasingly common with increasing redshift, 
going from of order a few hundred examples over the whole sky at $z<0.1$ to 
several hundred per square degree at $z\geq 1$. These distant LIRGs seem 
similar, at least superficially, to their low redshift  counterparts 
\citep[e.g.][]{far2,far3,cha03,sma03,tak06,val07}.
Excellent reviews of the properties of LIRGs can be found in 
\citet{san96}, and more recently in \citet{lfs06}. 

That the luminous end of LIRGs were discovered to be almost 
invariably mergers containing obscured starbursts and AGN linked them 
to several important galaxy transformational processes, but also raised 
many questions. Numerical simulations suggested that 
mergers could serve to transform disk galaxies into elliptical galaxies 
\citep[e.g.][]{bar90}, potentially linking LIRG activity to a key step in the 
formation of the familiar galaxy morphologies seen locally, but some recent 
simulations have found evidence that this may not always be the case
\citep[e.g.][]{bou05,rob06}.
The exact number of merger progenitors is also controversial, 
with some studies finding that LIRGs are mostly mergers between two galaxies 
\citep{veil02}, and others finding evidence for multiple mergers in a 
significant number of LIRGs \citep{borne00,far1}. The starbursts and AGN in 
LIRGs are obviously candidates for building up the stellar masses and large 
central black holes seen in ellipticals, but exactly how and when starburst 
and AGN activity are triggered as a function of merger stage, how the
star formation is distributed spatially and evolves in time, and how the 
starburst and AGN contributions to the total IR emission can be disentangled, 
are still debated. Two examples include the longstanding 
debate over the links between LIRGs and QSOs \citep{kaw06}, and the recent 
suggestion that the far-IR emission from LIRGs is determined more by whether 
the starburst or AGN is the dominant feedback mechanism, rather than which 
is more intrinsically luminous \citep{cha07}. The issue of feedback in LIRGs, 
in particular large-scale outflows, is also important for determining 
how LIRGs may contribute to the enrichment of the intergalactic medium
\citep{vei03,lip05}. 

The central position that LIRGs seem to play in 
galaxy evolutionary processes at all redshifts makes 
their in-depth understanding of great interest.
When considering studies of local LIRGs there are, broadly, two approaches; 
population studies of large samples to assess trends, and concentrated studies 
of single objects using panchromatic datasets. In this paper we adopt the 
latter approach, utilizing a comprehensive library of 
imaging and spectroscopic data to study the dynamics and power source 
within a nearby gas-rich interacting LIRG, to understand how
merger dynamics are affecting, and have affected, its starburst and 
AGN activity. Our dataset comprises imaging from 
the Advanced Camera for Surveys onboard the Hubble Space Telescope (ACS/HST), 
deep optical spectroscopy from the newly commissioned Southern African Large 
Telescope (SALT), adaptive optics near-infrared (NIR) imaging from the 
Very Large Telescope (VLT), and mid-infrared photometry and spectroscopy from 
the instruments onboard {\em Spitzer}.
Using this dataset we (1) construct a detailed 
picture of the merger dynamics in this system, (2) estimate the number of 
merger progenitors, (3) deduce the luminosities and other relevant parameters 
for starburst and AGN activity, (4) search for evidence for outflows, and
(5) discuss scenarios of the system's evolution. 
We assume a spatially flat cosmology with $H_{0}=73$ km s$^{-1}$ Mpc$^{-1}$, 
$\Omega=1$, and $\Omega_{m}=0.27$.

\section{Observations and data reduction}

\subsection{The target}

The luminous infrared galaxy IRAS 19115-2124, also known as ESO 593-IG 008, 
lies at a redshift $z=0.049$. It has an IR luminosity 
of $10^{11.9}{\rm L}_{\odot}$ \citep{sanders}, making it fall just short of a
formal ULIRG classification of $10^{12}{\rm L}_{\odot}$, 
and a molecular hydrogen mass of $3\times10^{10}{\rm M}_{\odot}$ 
\citep{mirabel90}. 
It is a classical example of a nearby, IR-luminous, gas-rich interacting 
system. The Virgo infall corrected distance to IRAS 19115-2124 with our 
adopted systemic velocity of $14576\pm9$ km~s$^{-1}$ is 200.0 Mpc, 
and the angular scale 0.97 kpc~arcsec$^{-1}$.

\subsection{NACO near-infrared imaging}

Near-infrared imaging of IRAS 19115-2124 was carried out  
using the NAOS-CONICA (NACO) adaptive optics (AO)
instrument on the VLT UT4.  The ${\rm K_{S}}$-band 
(the subscript will be dropped henceforth for simplicity) data set, 
using the S27 camera giving a field of view (FOV) 
of 27 arcsec and pixel size of 0.027 arcsec, 
presented here was taken in service mode on April 13, 2004.  
The visual wavefront sensor
in mode VIS-WFS 2-7 using a $V=12.9$ mag natural guide star 22 arcsec 
southwest of the target
was used for the AO correction.  Coherent energies of the reference star  
in the range  48 to 54 per cent were achieved in individual frames, while a 
strehl ratio (SR) $\approx0.25$ was measured
at the location of the target galaxy in the final image.  
The FWHM of point sources close to the target galaxy in the
final image are approximately 0.10 arcsec. The NACO $K$-band observations 
were repeated on September, 10, 2004, to allow searching for supernovae 
(SNe) within the nuclear regions of the
galaxy. 
This time coherent energies for the reference star ranged
between 36 and 51 per cent in the individual frames, yielding 
SR$\approx$0.18 at the location of the target galaxy in the final image. 

\begin{figure*}
   \centering
   \includegraphics[width=15.7cm]{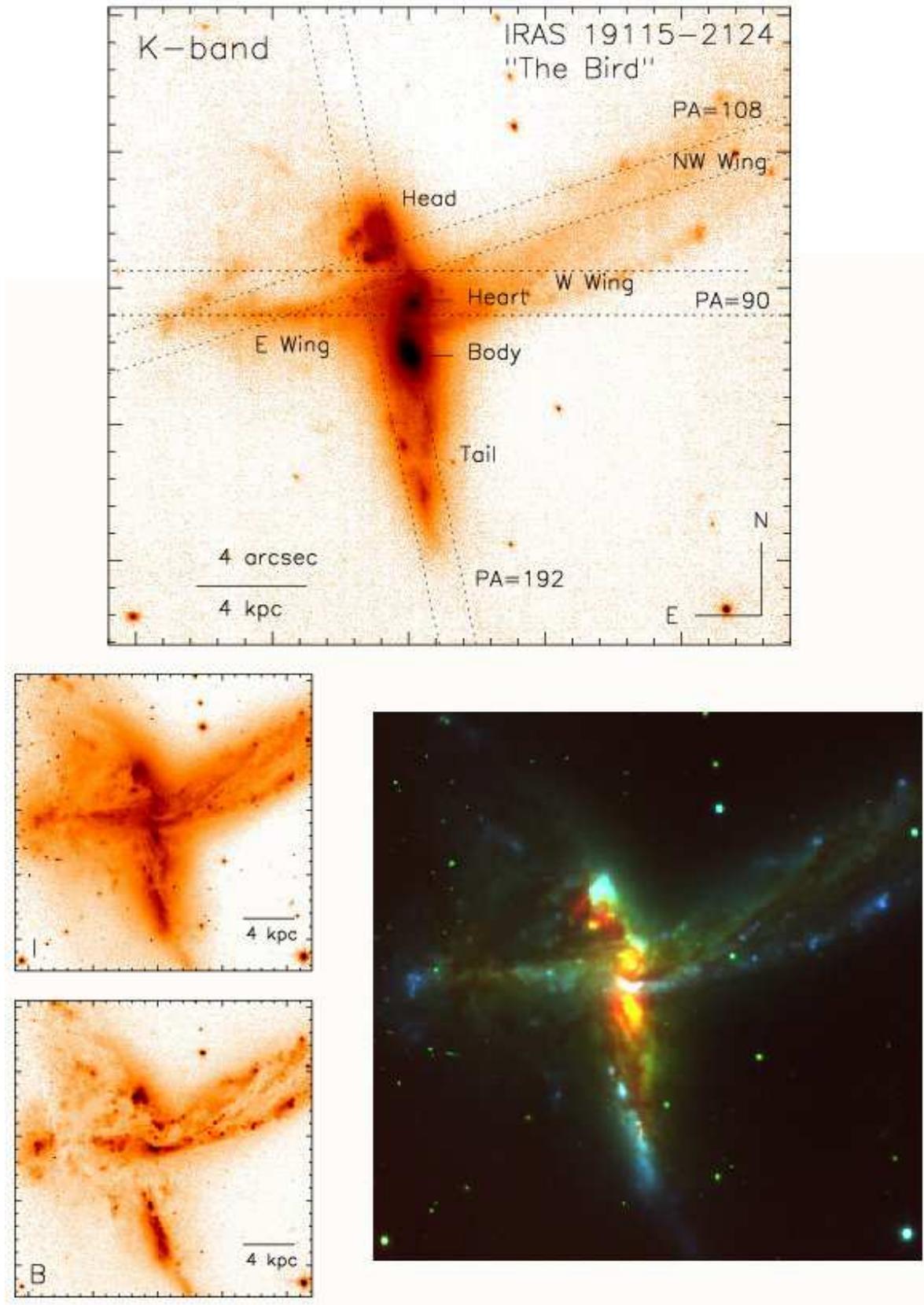}
   \caption{The NACO image of IRAS 19115-2124 (obtained on April 13, 2004) 
            is at the top, and the three slit positions observed
            with SALT/RSS are indicated, as are the main components
            of the interacting system, discussed and named in 
            Section~\ref{photometry}. At lower left are the HST $B$ and $I$
            images, and at lower right the combined $BIK$ 3-colour 
            image.
            All tick marks are in one arcsec intervals,
            and the brightness 
            scales are logarithmic, except for the 3-colour image.
            }
    \label{bigima}
   \end{figure*}

The (auto)jittered frames were median combined to 
form a sky image using IRAF. 
The sky subtracted 
images were then de-dithered using the centroid of a bright field star.
The AO correction was sufficiently stable over the whole observation
that 23 out of the 24 frames obtained on April 13 
were used to median combine the 
final image.  The individual frames had 90 sec exposure times, giving a 
total integration time of 2070 seconds for the final image of IRAS 19115-2124
(Fig.~\ref{bigima}). For the second epoch image all 24 of the 90 sec exposures 
taken were median combined to form the final frame with 2160 sec total 
integration time. 

We tied the photometry of the final image to the 2MASS $K$-band 
using large-aperture
photometry of the galaxy itself, as well as 2MASS stars in the FOV.
The resulting zero-point is also consistent with the ESO provided photometric 
standard star calibration, and we estimate the absolute Vega-based magnitude
to be accurate to within $\pm0.07$ mag.  The astrometric solution was found 
using GSC2 field stars. Four were available in the FOV, and the rms of 
the fit was 0.09 arcsec; the absolute GSC2 system has an uncertainty 
in the range 0.2--0.5 arcsec for individual stars.

\subsection{Optical observations}

\subsubsection{HST/ACS imaging}

We extracted archival ACS images (PI: Evans) 
of the Bird in F435W and F814W bands, i.e.\ in the HST $B$ and $I$-bands, 
of exposure times 1260 and 720 seconds, respectively. 
These data are already calibrated
and we used the supplied Vega-based zeropoints.  We did not perform 
any additional processing of the images, except a flux-conserving geometric 
transformation into the NACO image size and resolution for the purpose
of photometry in identical apertures. The images, as well as a 3-colour 
image with the $K$-band included, are shown in Fig.~\ref{bigima}.

\subsubsection{SALT/RSS spectroscopy}
\label{rssspectra}

Spectroscopic observations presented here were obtained with the Robert 
Stobie Spectrograph \citep[RSS,][]{burgh03,kobul03}  
during the commissioning and performance verification stage
of SALT \citep{buckley06,dod06}, a new 10-m class telescope in Sutherland, 
South Africa.

The RSS observations are described in Table~\ref{table1}.  We placed slits 
through the galaxy system in three different position angles (PA) on three
different nights. The slit positions are indicated in Fig.\ref{bigima}.  
The same grating (PG1800) with a dispersion of 
0.41 \AA$ \ {\rm pix}^{-1}$ and spectral resolution of 2.4~\AA\ FWHM (with
1.5 arcsec  slit), and spectral range from $\approx5860$~\AA\ to 
$\approx7150$~\AA\ was used each time.
The October 22 data set is of the best quality, 
taken in both good seeing and transparency conditions.
The October 10 observations were taken through thin clouds, and have lower
signal to noise ratio (SNR) than the others, 
though all the strong emission lines are very clearly seen.
Spectra of Cu-Ar and Ne comparison lamps were taken after the science frames.

The data were first bias and overscan subtracted, gain corrected, trimmed,
cross-talk corrected, and mosaiced, using a specially adapted {\it salt} IRAF
package, and cosmic ray corrected using MIDAS routines.  The IRAF 
{\it twodspec} package was used to wavelength calibrate and correct each 
frame for distortion and tilt, and to background subtract the 2D spectra.
The {\it apall} routine was used to extract various 1D spectra.
The derived internal errors for the 2D wavelength calibrations are 
$\sigma=0.05$~\AA\ throughout the wavelength range of detected emission lines, 
i.e. $<2.5$ km~s$^{-1}$, except for September 19 
where $\sigma=0.09$~\AA. In addition,
a check of night sky lines throughout the wavelength range,
using methods described in \citet{zasov}, confirmed the
accuracy of the calibration to this level.  All line-of-sight velocities
presented here were then corrected for heliocentric motion.

\begin{figure*}
   \includegraphics{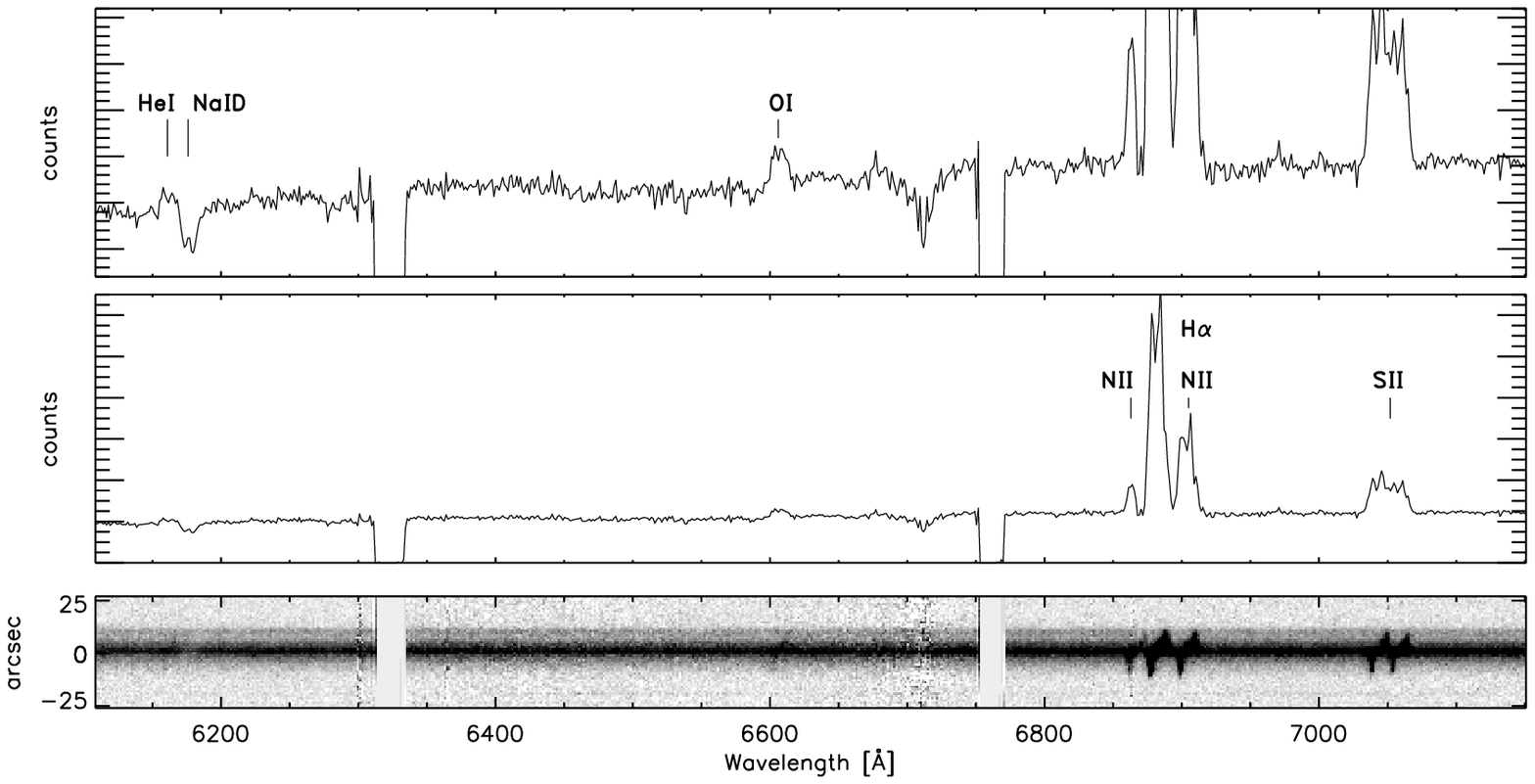}
   \caption{The reduced 2D spectrum around the target 
          at PA=90 covering nearly the 
          whole observed wavelength range (cut slightly in the blue).
          The 1D spectrum is extracted from within a wide 10 arcsec 
          aperture.
          The two gaps in the spectrum are gaps in between the 
          3 CCD chips of the RSS detector, and the high noise area
          at $\approx 6720$~\AA\ is an incomplete subtraction of scattered
          light from SALT 'autocollimator laser'.}
   \label{figfull2d}
   \end{figure*}

\begin{figure}
   \includegraphics[width=84mm]{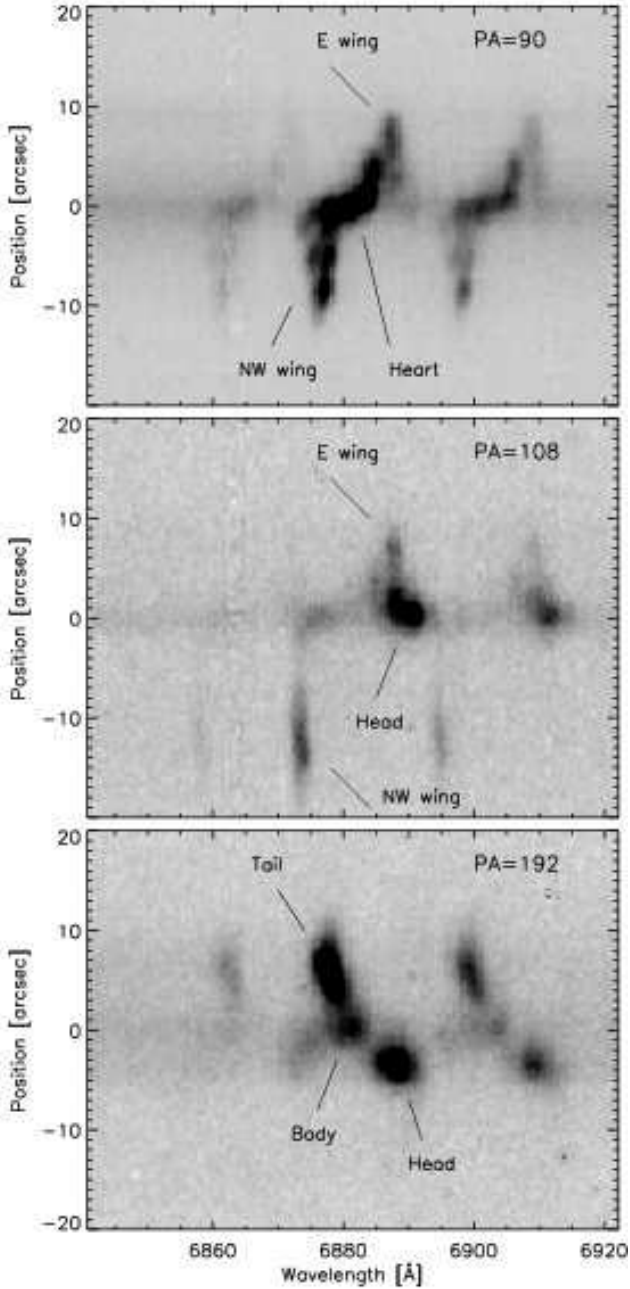}
   \caption{The reduced 2D spectrum around the H$\alpha$ and \ion{N}{II} lines
            from three different position angles. The main components of the 
            merging system are indicated, see Fig.~\ref{bigima}.  
            The bright emission lines can be traced for over 20 arcsec,
            i.e.\ 20 kpc.}
   \label{figha}
   \end{figure}

Figure~\ref{figfull2d} shows the 1D and 2D spectra of the PA=90 data, 
while Fig.~\ref{figha} shows the 2D spectral region around 
H$\alpha$ of all three slit positions. 

\begin{table}
 \centering
 \begin{minipage}{84mm}
  \caption{Log of VLT NIR imaging and SALT optical spectroscopy 
           of IRAS 19115-2124.}
  \label{table1}
  \begin{tabular}{l c r c r}
  \hline
          &  Date      & Exp.time & Band   &  \\  
 \hline
VLT/NACO  & 13/04/2004 & 2070 sec    & $K_s$ & \\
VLT/NACO  & 10/09/2004 & 2160 sec    & $K_s$ & \\
 \hline
          &  Date & Exp.time & Slit   & PA \\  
 \hline
SALT/RSS  & 19/09/2006 & 1500 sec &  1.5\arcsec & 192 deg \\ 
SALT/RSS  & 10/10/2006 &  900 sec &  1.0\arcsec & 108 deg \\ 
SALT/RSS  & 22/10/2006 & 1200 sec &  1.5\arcsec & 90 deg \\ 
\hline                          
\end{tabular}
\end{minipage}
\end{table}

\subsection{{\em Spitzer} MIR imaging and spectroscopy}
\label{spitzer}

\begin{figure*}
   \centering
   \includegraphics{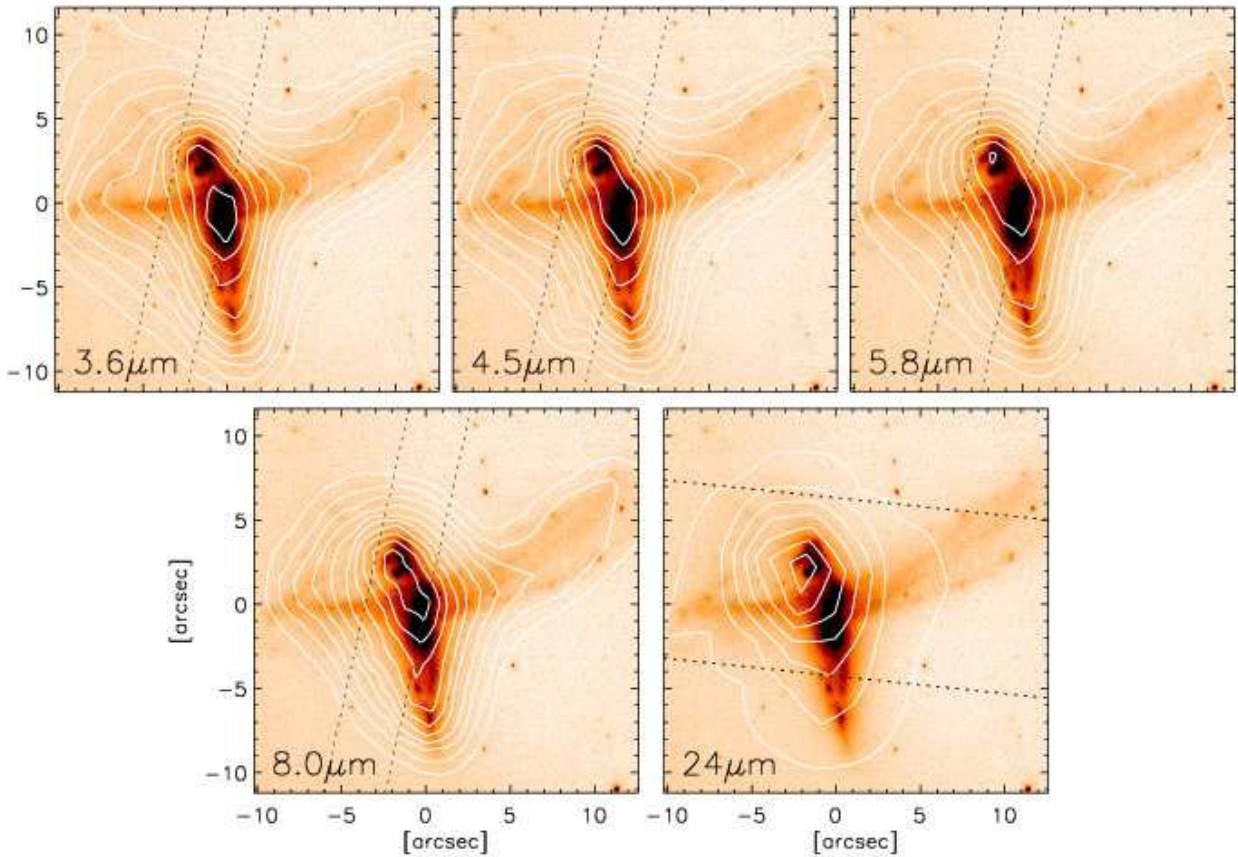}
   \caption{{\em Spitzer} IRAC 3.6, 4.5, 5.8, and 8.0 $\umu$m, and
            MIPS 24~$\umu$m contours are overlaid on the NACO $K$-band data.  
            Contour levels are logarithmic.
            The peak flux from {\em Spitzer} bands shifts 
            from the Body towards the 
            Head NIR component as the wavelength increases. The dotted lines
            on IRAC panels denote the width and orientation of the slit used 
            in IRS LS spectrum (5.2--14.5 $\umu$m), and the MIPS panel is 
            overplotted with the LL slit (14--38 $\umu$m).}
              \label{spitzerall}
    \end{figure*}

We obtained archived {\em Spitzer} IRAC (3.6, 4.5, 5.8, and 8$\umu$m)
and MIPS (24$\umu$m) images of IRAS 19115-2124 (PI: Mazzarella).
We used the post-basic calibrated data (PBCD) 
products provided by the S14.0.0 and
S14.4.0 versions of the pipeline for the IRAC and MIPS data, respectively.
The IRAC PBCD mosaics were rebinned to the pixel scale of NACO and
were aligned to the same orientation and pixel coordinates using
centroid positions of 2--4 point sources visible in both images. 
Fig.~\ref{spitzerall} shows the IRAC and MIPS contours.

\begin{figure*}
   \centering
   \includegraphics{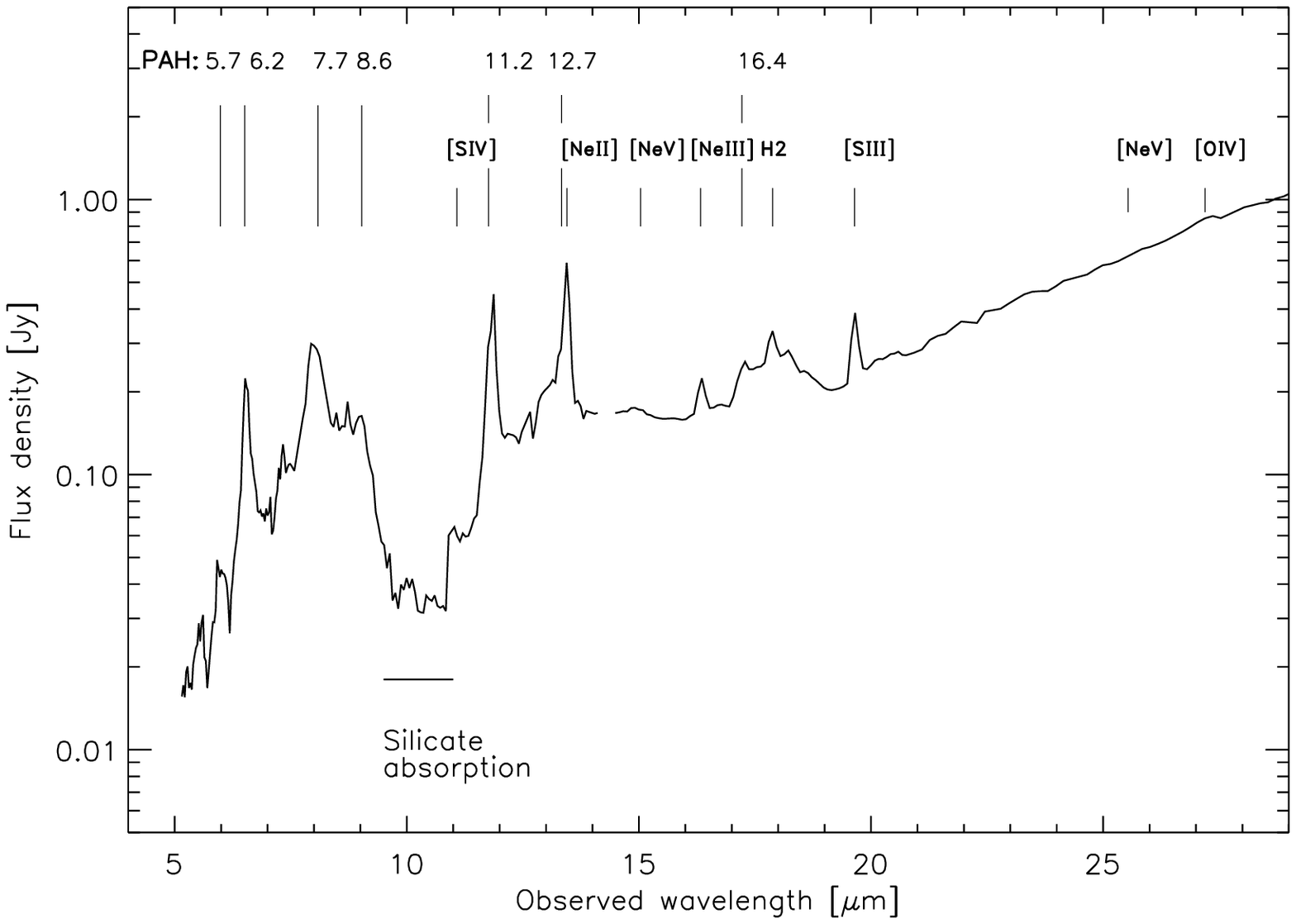}
   \caption{A {\em Spitzer}/IRS spectrum of the Bird. 
            The main PAH features and other detected emission lines are 
            indicated.  The spectrum continues featureless 
            beyond the plotted area. The small discontinuity at $\sim14$
            $\umu$m is the transition from SL to LL mode observation.}
   \label{irs}
    \end{figure*}

{\em Spitzer} IRS data (PI: Armus) were also extracted from the archive. 
The data consists of SL (5.2 -- 14.5 $\umu$m 
wavelength range, 3.7 arcsec slit width) and LL-mode 
(14 -- 38 $\umu$m, 10.5 arcsec slit width). 
Figure~\ref{irs} shows the spectrum with the most important
polycyclic aromatic hydrocarbon
(PAH) and emission lines indicated. The orientations of the SL and LL mode
slits are overlaid on the IRAC and MIPS contour images, respectively,
in Fig.~\ref{spitzerall}. 

The IRS data were reduced using the SMART and SPICE software packages.  
The individual BCD frames for each nod position were cleaned of cosmic  
rays and other artifacts using the {\em irsclean} software package, 
and  combined 
into a single image using the `fair coadd' option within  SMART. Sky 
subtraction was performed for each module by using the  off-order image 
from the same module (so called 
'order-order' sky  subtraction). Spectra were then 
extracted and flux-calibrated from these sky-subtracted, combined images 
using the 'optimal extraction'  option within SPICE. All other parameters, 
such as extraction apertures ($>14$ arcsec), 
were left at their default values. The 1D 
spectra for each  nod position were then combined to give the final spectrum.
However, since the IRS slits do not completely
cover the galaxy system, and because of the varying slit size in the
short and long wavelength part of the IRS spectrum, we multiplied the
SL section of the this final spectrum by a factor of 1.6 and the LL section by 
1.2.  These corrections were estimated from flux falling outside the slits
in the IRAC and MIPS maps, and we estimate them to be accurate to 
$\sim$10--15 per cent.  The IRS spectrum agrees with broad band
photometry after the correction. 

\section{Analysis}

\subsection{System characteristics from imaging}

\subsubsection{General morphology and photometry}
\label{photometry}

On the basis of the $K$-band NACO AO-image alone, IRAS 19115-2124 appears
to be a major merger system with at least two prominent components, 
perhaps three, with tidal
tails in addition (see Fig.~\ref{bigima}).
With its outspread wings, a head complete with a beak, 
a body and a tail, we dub this merging galaxy system the Bird.

We present photometry of the major components in the optical and NIR
bands below. For these SExtractor \citep{bertin} 
and IDL based aperture photometry was used, and all the 
magnitudes are Vega based.  We note that
no aperture corrections, or corrections dependent on the slightly varying SR
ratio over the image are needed when the aperture used is greater than 
$0.5$ arcsec in radius, or in the case of extended objects. 
Photometry of smaller point-like sources is explained in Sections~\ref{ssc}
and~\ref{snsearch}.

The largest and brightest component in $K$-band is the Southern galaxy 
nucleus, or the Body of the Bird.  It is severely
extinguished and virtually invisible in the ACS $B$-band image.
The bright NIR extent of this galaxy is $\sim2$ kpc.  
In a 1 kpc radius aperture (1.03 arcsec) the apparent
brightness of the central region is $K$=12.82 mag, corresponding to 
$M_K = -23.69$ absolute magnitude.
This is not much fainter than an $L^{\star}$ galaxy, $M^{\star}_{K}=-24.1$ 
\citet{cole01}.  Magnitudes in other apertures and all bands are
tabulated in Table~\ref{phottable}.
The galaxy is quite disturbed and extends 
South with a more than 6 kpc long Tail of the Bird, which has several clear
non-pointlike, elongated 0.3--1.0 kpc sized subconcentrations. 
The brightest two of these have $K\approx 17.0$ ($B=20.9$, $I=20.0$ mag), 
i.e.\ $M_K = -19.5$ inside 300 pc radii. 

The Heart of the Bird consists of a fairly regular looking barred spiral
(in $K$-band), with a wide opening of the spiral arms.  The arms can be
traced out to at least 2 kpc from the nucleus. In a radial aperture of 1 kpc
the brightness is $K$=13.35 mag, ie. $M_K = -23.16$. In the optical, the
spiral arms are more pronounced than the nucleus, though only small 
sections of the arms can be seen in the $B$-band.

The Head, the northernmost component, appears to be an irregular galaxy, 
extending
approximately 2 kpc with numerous distinctive bright knots in the $K$ image, 
presumably bright \ion{H}{II} regions and super star clusters (SSC). 
The optically brightest section is 
the northern part of the Head, while the brightest NIR knot 
is in the most obscured southern part of the Head. 
The total brightness of the Head is 
$K$=13.80 mag in a 1 kpc radial aperture ($M_K = -22.71$) i.e. only 
half a magnitude fainter than the Heart. 
Individual knots, excluding the more diffuse K-band background light, 
range in brightness from $K=17.6\pm0.1$ to $K\approx20.0$.  
We can isolate 12 of these \ion{H}{II} regions mainly at the 
edge of the component closest to the Heart; the 
blobs get fainter with increasing distance from the Heart and Body.
Most of these blobs are in highly extincted regions and are not visible
in the optical images -- an elaboration on point-like SSC 
candidates is presented below in Section~\ref{ssc}.

Additionally, there are tidal tails, the Wings, extending East  
and North-West, visible for 8 kpc and 15 kpc, respectively.
In addition we separate a West Wing, a shorter
section of the tidal tail below the larger structure extending North-West,
and which appears from the ACS images to continue from the Heart spiral
arms.  All the Wings also have numerous bright knots 
though not as 
extended as the ones in the Tail.  Finally, there is a very complex
dusty area North of the E-wing and to the NE of the Head.

\begin{table}
 \centering
 \begin{minipage}{84mm}
  \caption{Photometry of IRAS 19115-2124. Absolute magnitudes can be
           calculated with a distance modulus of 36.51.}
 \label{phottable}
  \begin{tabular}{l c c c c} 
  \hline  
Component &  aperture\footnote{The 1.5, 1.0 and 0.5 kpc radius apertures are
calculated above the global sky level, the 0.2 kpc aperture is above the local
galaxy background. Photometric errors in all bands are dominated by 
systematic effects and are approximately at 0.05 mag level --
only the 0.2 kpc aperture $B$-band values have errors of the order of 
0.2--0.4 mag.}     &  $B$    & $I$     &  $K$    \\
          &  (kpc)     & (mag) &(mag)  & (mag) \\
  \hline
Bird total&  15\footnote{Flux within a large 30 arcsec diameter.  
$K$-band value is also equal to the 2MASS total magnitude} &  15.6   &  13.6  & 11.0 \\
Body      &  1.0       & 19.80   & 17.00  & 12.82 \\
...       &  0.5       & 22.00   & 18.00  & 13.58\\
...       &  0.2       & $>25.7$ & 19.32  & 15.68 \\
Heart     &  1.0       & 19.31   & 16.87  & 13.35 \\
...       &  0.5       & 21.52   & 18.22  & 14.41 \\
...       &  0.2       & 23.81   & 19.63  & 16.63\\
Head      &  1.5       & 18.64   & 16.49  & 13.39 \\
...       &  1.0       & 21.51   & 17.13  & 13.80 \\
 \hline
\end{tabular}
\end{minipage}
\end{table}

\subsubsection{Surface brightness profiles and effective radii}

To study the morphologies of the galaxy 
components of the Bird more quantitatively,
we used the GALFIT \citep{peng02} software on the NACO data 
to fit 2D surface brightness distributions to the two brightest 
components, the Heart and the 
Body, and in addition fit simple surface brightness profiles 
to cuts along the semi-major axes of these two galaxies. All the systems
of the Bird are quite disturbed, so unique solutions, especially with
GALFIT, are difficult to find.  Nevertheless, results are 
satisfactory for the purpose of classifying the galaxies, and to derive 
effective radii to better than a factor of 2 accuracy, that is needed in 
mass estimates later.

\begin{figure*}
   \centering
   \includegraphics{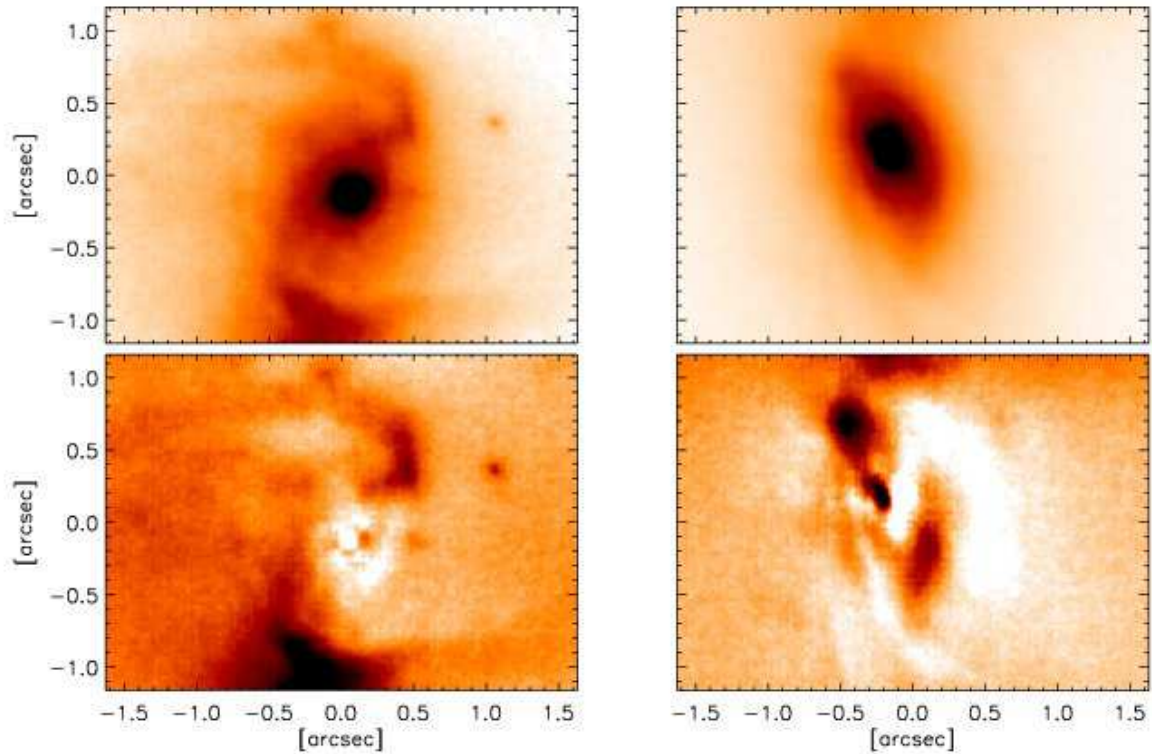}
   \caption{Close-ups of Heart (left) and Body (right) galaxy nuclei of
            the Bird from the $K$-band AO data.  Top panels
            show the data, and the bottom panels shows the respective 
            galaxy nuclei after subtracting the GALFIT
            best-fitting bulge and disk components -- see text and 
            Table~\ref{galfitpar} for the values.  The brightest residuals 
            are of the order of $\sim7$ per cent of peak fluxes
            for both nuclei. There is a bar-like 
            structure revealed in the Body nuclear region, 
            which was not obvious in the unsubtracted image. Images are
            shown with inverted brightess scale.} 
           \label{figspirals_resid}
  \end{figure*}

\begin{figure}
   \centering
   \includegraphics[width=84mm]{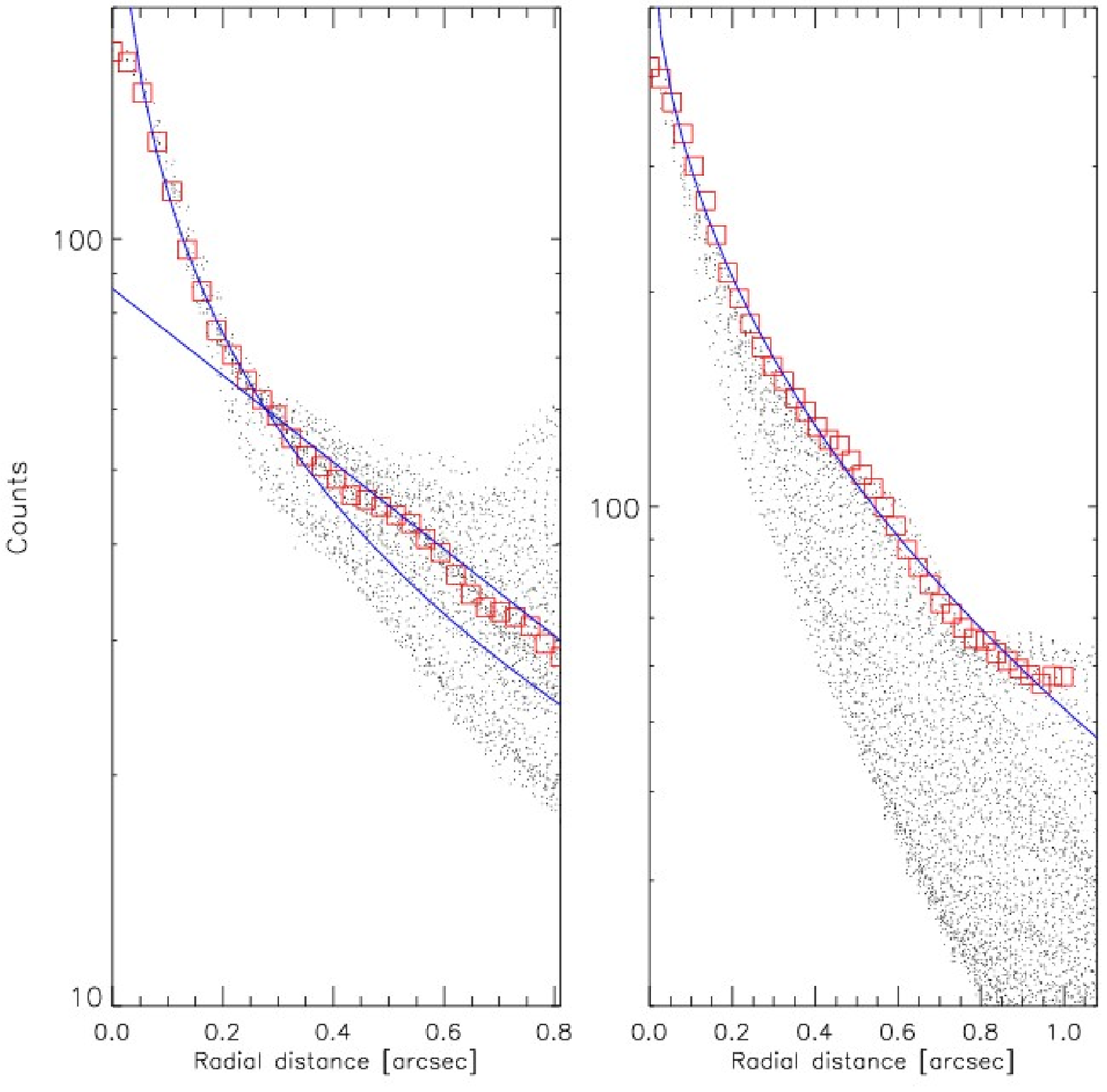}
   \caption{$K$-band surface brightness profiles of
            the smaller barred spiral (Heart, left), and 
            the larger NIR galaxy (Body, right).  
            The dots represent all pixel values
            as a function of distance from the nuclei, and red squares are
            values averaged over 0.1 arcsec bins along the major axis only.
            The dots above the squares in the left panel come from the bars.
            The blue lines are chi-square fitted profiles along the major axes:
            an exponential and de Vaucouleurs component in the
            left panel, and a single Sersic profile on the right,
            see Table~\ref{galfitpar} for best-fitting values.}
              \label{fig_profiles}
    \end{figure}

First of all, we do not find point-source components from the central
parts of the galaxy system.
The Heart of the Bird, the apparent barred spiral galaxy, is a
combination of a de Vaucouleurs bulge component and an exponential disk.  
The best-fitting effective radius for the disk
component is $r_e$=1.2 kpc using a 1D profile along the semi-major axis 
(see Fig.~\ref{fig_profiles}), and when fit out to a distance of 1 arcsec.  
GALFIT turned out to be quite sensitive to the adopted 'sky' value, i.e. 
in this case the value assigned to the underlying light from the tidal tails 
and other neighbouring components. With a realistic range of experiments,
GALFIT produces a range of $r_e = 1.1 - 2.3$ kpc, consistent
with the simpler method. Table~\ref{galfitpar} gives the best-fitting 
GALFIT parameters for all components.  The derived bulge-to-disk
ratio with these parameters is B/T = 0.09, i.e. typical of late type
spirals.
The lower left panel in  Fig.~\ref{figspirals_resid} shows the residual image 
after subtraction of both the bulge and disk components using GALFIT.

\begin{table*}
 \centering
 \begin{minipage}{140mm}
  \caption{GALFIT and 1D profile fit parameters for IRAS 19115-2124 NACO data.}
 \label{galfitpar}
  \begin{tabular}{l l l c c c c c c} 
  \hline  
Component & subcomp & Method &  $K$  & $r_e$  & $n$   & $b/a$  & PA  
          & boxiness\footnote{parameter $c>0$ boxy, $c<0$ disky} \\
          &        &         &(mag)& (kpc)  &       &        &(deg)&       \\ 
  \hline
Heart..   & bulge &  GALFIT\footnote{The errors reflect systematic 
        uncertainties of parameters found when using a range of fitting 
        areas, initial values etc. Formal GALFIT errors of parameters are 
        much smaller.} 
&  16.3$\pm$0.2        & 1.3$\pm$0.2   & 4 & 0.9$\pm$0.1 & -60$\pm$10 & -- \\
          & disk  &  GALFIT&  13.7$\pm$0.2        & 1.9$\pm$0.4   & 1 
          & 0.4$\pm$0.1 &  7$\pm$5 & -- \\
          & disk  &  1D\footnote{Formal errors are given for 1D fit 
                            parameters that were not fixed.}    
&  --                  & 1.26$\pm$0.03 & 1 & --   &  12 & -- \\
Body ...  & bulge & GALFIT &  16.1$\pm$0.2        & 0.7$\pm$0.2   
          & 2.3$\pm$0.2 & 0.7$\pm$0.1 & 33$\pm$10 & +0.6$\pm$0.1\\
          & disk  & GALFIT &  12.4$\pm$0.2        & 2.1$\pm$0.4   
          & 1.6$\pm$0.1 & 0.5$\pm$0.1 & 4$\pm$7 & +0.1$\pm$0.1\\ 
          & disk  &  1D    &  --                  & 2.52$\pm$0.05 
          & 2.23$\pm$0.04 & --   & 13 & -- \\ 
 \hline
\end{tabular}
\end{minipage}
\end{table*}

The Body of the Bird, a very disturbed galaxy, was more 
difficult to fit using GALFIT, and impossible with simple exponential 
and/or de Vaucoulers components. 
Using a 1D slice through the galaxy (Fig.~\ref{fig_profiles}) out to a 
distance of 1 arcsec, a single Sersic profile with
n=2.2 and $r_e = 2.5$~kpc is the best fit.  
Using GALFIT, a reasonable fit is found with two 
separate Sersic profiles with n=2.3 and n=1.6 for the nuclear and outer region
components, respectively.  Though the best fit here was again sensitive to 
the background value, and the index $n$ is coupled to the radius,
a reasonable range of parameters resulted in  
$r_e \approx 1.0 - 2.3$~kpc. 
The right panel of Fig.~\ref{figspirals_resid} shows the 
original surface brightness distribution, and the residual after the 
best-fitting subtraction, which
reveals a complicated underlying appearance, with evidence for
a bar-like structure extending NE-SW, and perhaps remains of spiral arms.
It is also noteworthy that regardless of the other parameters, the 
2D GALFIT profile for the nuclear region of the Body is always ``boxy'' 
(with the GALFIT boxyness parameter $c\approx +0.6$).

\subsubsection{Search for supernovae in NACO images}
\label{snsearch}

\begin{figure}
   \centering
   \includegraphics[width=65mm]{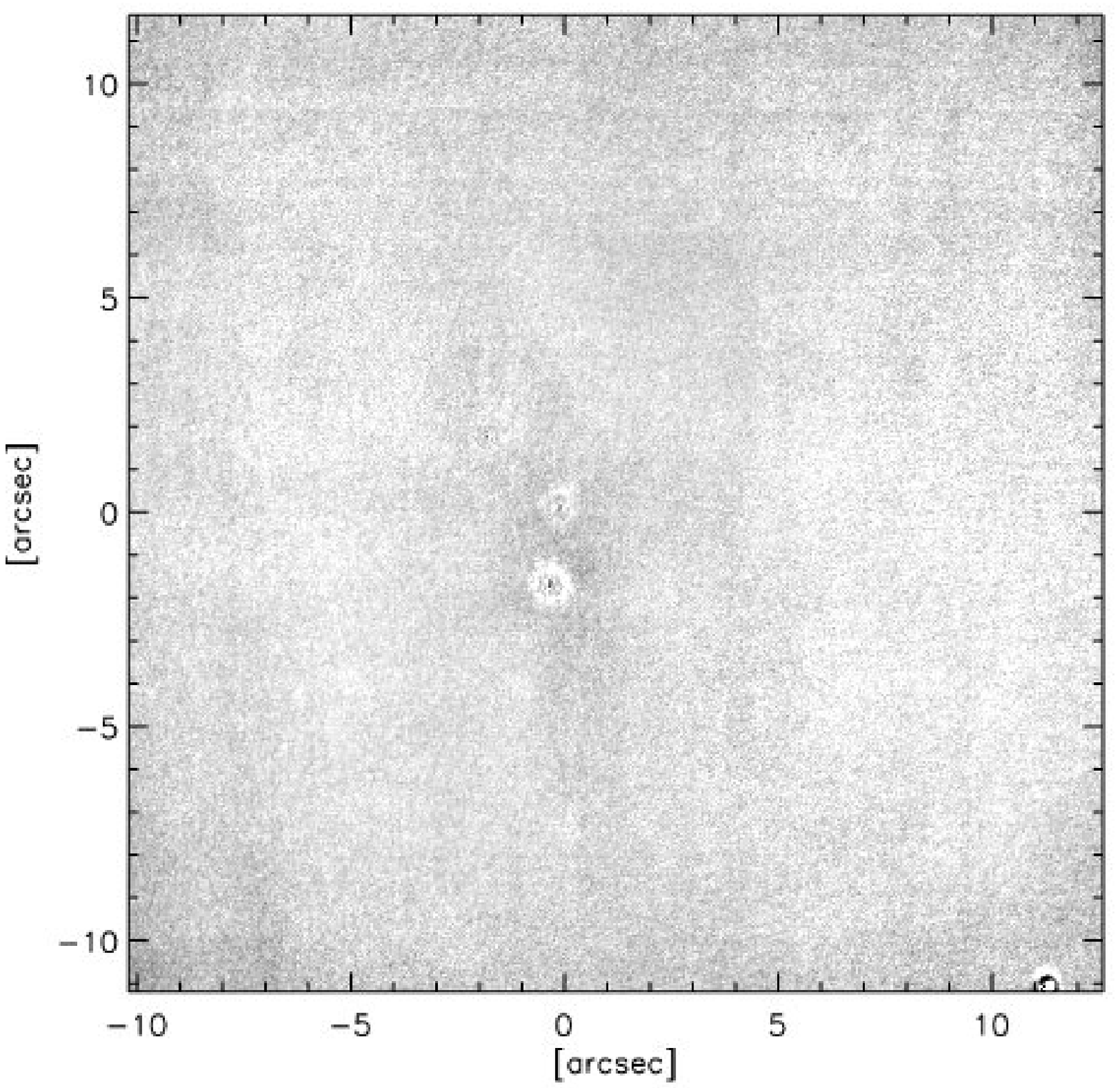}
   \caption{The two epochs of NACO $K$-band data subtracted from 
            each other (Sep 10 -- Apr 13).  No point-sources, SNe, 
            are found down to $K = 21.3$
            mag. The image shows mostly random
            noise but some significant residuals, due to differences in the
            seeing halos between the two images, are apparent around the
            location of the bright Body of the galaxy.  The brightness scale 
            is inverted and the image area
            is identical to those in Fig.~\ref{spitzerall}.}
           \label{snsub}
    \end{figure}

We have NACO $K$-band images obtained at two different epochs separated by
153 days. This allows us to search for supernovae that have exploded
within the nuclear regions of IRAS 19115-2124 during this period or before
the first epoch image. The images were aligned, subtracted and analysed in 
the manner described in \citet{seppo07}, where we detected the SN 2004ip 
in a similar NACO dataset of the LIRG IRAS 18293-3413.
No obvious point-sources were
found from the subtracted images (Fig.~\ref{snsub}). 
We then estimated SN detection limits by placing simulated
sources made from a real PSF of a nearby bright star
in a grid of locations within the nuclear regions of the Bird prior to 
the image subtraction.
We used an aperture radius of 0.15 arcsec and a sky annulus between 0.20 
and 0.30 arcsec for measuring both the noise and the simulated sources
in the subtracted images. 
This yielded $K = 21.5$ as the 3$\sigma$ limiting magnitude for the nuclear SN 
detection within the Head.  Within the Body, where larger image subtraction 
residuals were seen, we obtain a 3$\sigma$ limiting magnitude of $K = 21.3$. 
However, within the innermost $\sim$0.3~arcsec (or $\sim$300 pc) diameter of 
the Body the SN detection is more difficult and a SN of $K = 21$ mag 
would remain undetected in our NACO images. 

\subsubsection{Super star cluster candidates}
\label{ssc}

\begin{figure}
   \centering
   \includegraphics[width=84mm]{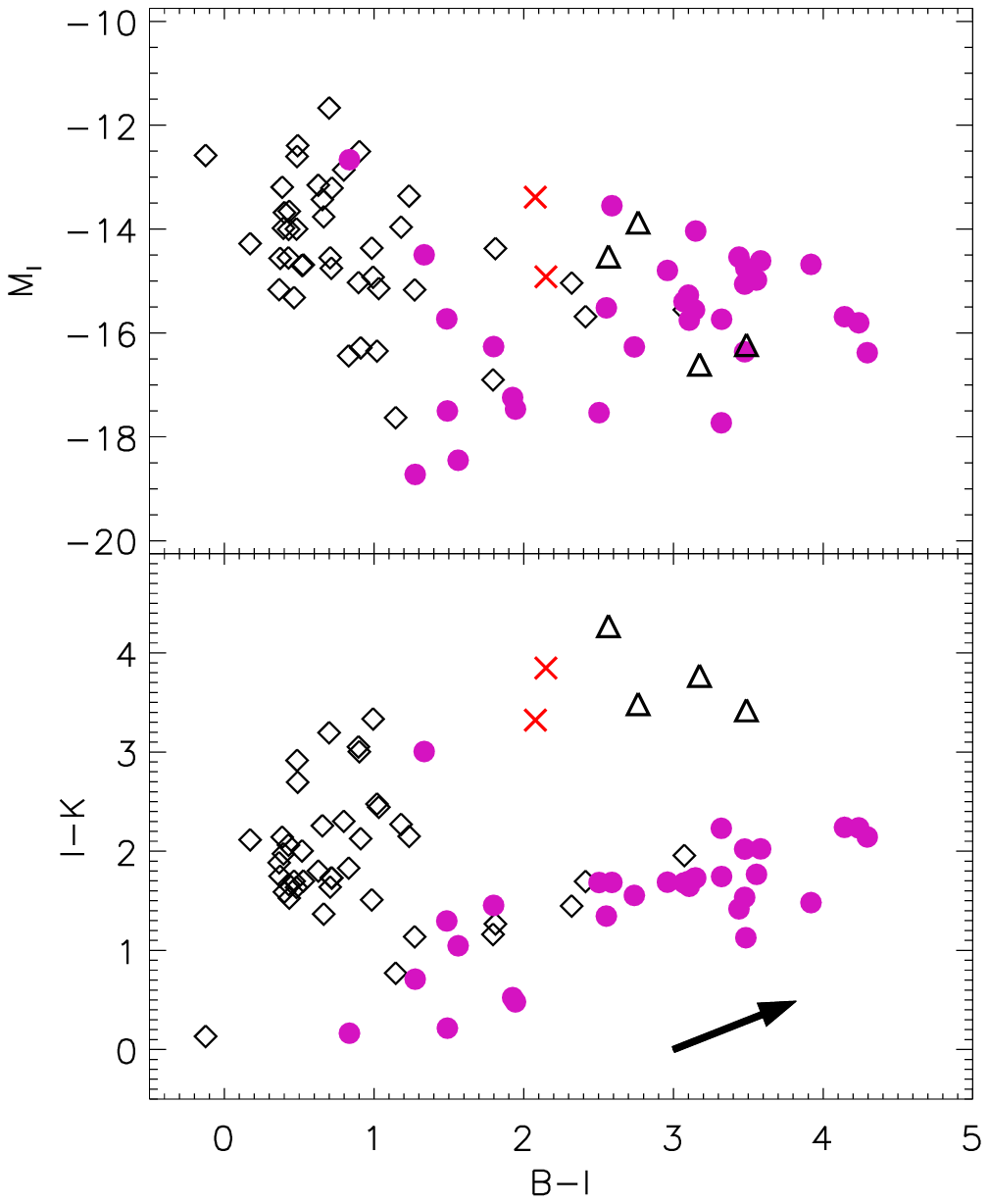}
   \caption{SSC candidates in the NACO image FOV, i.e.\ all point-sources 
            within the galaxy system (diamonds) and outside it (circles).
            Triangles show detections within core regions of the Head
            and Heart and crosses denote marginal NIR detections. 
            A reddening vector of $E(B-V)=0.32$ is indicated.
            }
           \label{sscbird}
    \end{figure}

To create a list of SSC candidates we searched for all point-sources in the 
ACS images within the NACO image FOV -- 83 were found -- and aperture
photometry was performed in 0.2 arcsec radii apertures and the 
background measured in the surrounding 0.1 arcsec sky annuli. 
Aperture corrections for each ACS image  
were applied \citep{sirianni}, and Galactic extinction was corrected for.
Identical apertures were then used on the NACO image, and 78 of the 
83 point-sources were detected, 2 of them marginally
(cross points in Figure~\ref{sscbird}).  Aperture corrections 
of 0.84 mag were applied, determined from brighter isolated point-sources 
in the field; scatter is of the order 0.1 mag, and a small dependance on the
distance to the WFS reference star was ignored, since it is smaller than the 
scatter.
 
Figure~\ref{sscbird} shows plots of the SSC candidates, the
different symbols differentiate between point-sources inside the Bird system, 
(45 of them, open diamonds), and those outside the system (31, filled circles).
The largest single concentration of these point-sources is along the
W and NW Wings.
The number density of the measured sources inside and outside
of the galaxy was compared to the number density of point sources
(foreground stars) in the ACS images outside the NACO FOV.
From this we estimate that $\sim10$ per cent of the SSC candidates
inside the Bird and $\sim90$ per cent of those
outside the system are foreground stars.
Therefore, the colour-colour diagram (Fig.~\ref{sscbird}, lower panel)
appears to nicely separate the SSCs from the foregound stars. 
A handful of points have very red $B-I\sim3$ and $I-K\sim4$ colours;
these are all 
within the core regions of the Head and Heart.
We note that based on Fig.~\ref{sscbird} the luminosity function of 
the SSC candidates
appears to be brighter by at least a magnitude than e.g.\  in the local
luminous interacting galaxy prototype Antennae \citep{anders07}.
However, we defer more detailed discussion on SSC candidates in the Bird 
to a separate work. 

\subsubsection{MIR emission}

Figure~\ref{spitzerall} shows the Bird $K$-band data
overlaid with {\em Spitzer} IRAC and 24~$\umu$m MIPS contours.
The 3.6 $\umu$m data closely
matches the $K$-band, but the longer the wavelength, the more
of the relative flux comes from the northernmost Head component, 
the irregular galaxy with numerous \ion{H}{II}/SSC regions.  
We also used the 3.6 $\umu$m band map 
to extrapolate an approximate stellar contribution
in the other IRAC bands by simply assuming a Vega-like
stellar spectrum \citep[see e.g.][]{pahre04}. 
The stellar light subtracted 5.8 and 8.0 $\umu$m maps do not differ 
appreciably from the non-subtracted ones, which is 
not surprising since these are expected to be
dominated by non-stellar emission everywhere. The stellar-subtracted
4.5~$\umu$m image, however, clearly reveals the Head to be the major
source of non-stellar MIR flux.

\begin{figure}
   \centering
   \includegraphics[width=84mm]{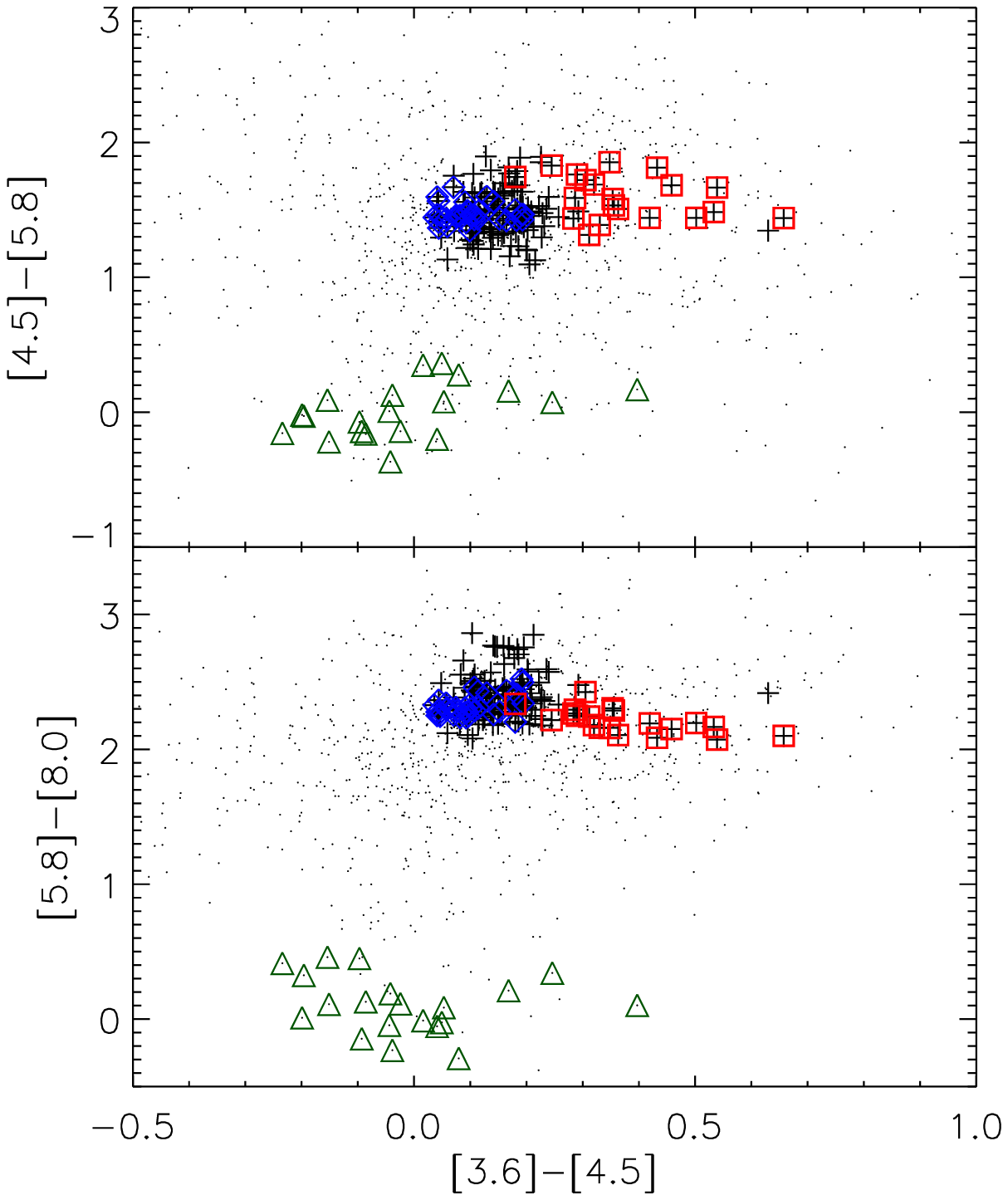}
   \caption{Pixel-by-pixel IRAC 
            colours of IRAS 19115-2124 area.
            Black crosses are pixels within the
            the galaxies (defined as $f_{8.0} > 6$ MJy/sr), and the green 
            triangles are the two bright stars in the field.  The blue
            diamonds pick out the area of the NW wing, and the
            red squares the Head of the Bird -- the latter 
            is clearly separated with the [3.6]-[4.5] colour.
            The tidal tails, in contrast, 
            do not differ from the bulk of the system.
            }
              \label{figcolcol}
    \end{figure}

MIR colours are examined in more detail in Figure~\ref{figcolcol},
where all IRAC pixels in a 22 by 22 arcmin area around the Bird are
plotted.  Pixels at the locations of two field stars are located
close to zero-colours as expected.  Most colours of the
interacting system are close to those expected from late type spiral and
irregular galaxies, e.g.\ [3.6]-[4.5]$\sim$0.1 and [4.5]-[5.8]$\sim$1.5
\citep{davoodi06,smith07}.
There are two significant exceptions, however. All colours with the  
8.0 $\umu$m IRAC band are very red, indicating stronger than normal spiral 
galaxy PAH emission contribution in that band.  Secondly, the Head
(red squares) has very red colours  
in [3.6]-[4.5] and [3.6]-[5.8], more typical of e.g.\ Seyfert nuclei. 
This must be a result of warm to hot dust in the irregular galaxy,
which can in principle be heated either by very strong star formation, or AGN
activity. We investigate spectral evidence for AGN further in 
Section~\ref{lines}.

\subsection{Analysis of spectra}

\subsubsection{Physical characteristics from emission lines}
\label{lines}

Several warm gas emission lines are readily detected in the SALT/RSS
spectra, as seen for example in Fig.~\ref{figfull2d}.
Table~\ref{components} summarizes the optical line ratios available from
our data for the different parts of the Bird, gathered from all three
slit positions, and calculated using multi-component fitting with the 
IRAF task {\em splot}.
The line ratios are not corrected for 
stellar absorption; with typical EW$(H\alpha) > 50$~\AA\ (see below) 
we expect the correction to be not significant. 
All areas have uniform ratios, in the range
[\ion{N}{II}]$\lambda$6583/H$\alpha \sim 0.40-0.45$ and 
[\ion{S}{II}]$\lambda\lambda$6716,6731/H$\alpha \sim 0.4$ both suggesting 
\ion{H}{II} type 
regions \citep{veil87}.  
The only notable exceptions are the nuclear region of the Heart, 
and the deblended blue wing from the Heart: while the nucleus has even more 
solid \ion{H}{II} type ratios than the average,  
the blueshifted emission from the Heart has an elevated [\ion{N}{II}] level 
with the ratio to H$\alpha$ at $\sim0.6$ (similar to other deblended 
blue-shifted components as well, which are not tabulated).  
This is close to the \ion{H}{II} vs.\ AGN 
separation line; but instead of photoionization by an AGN, it may well 
indicate shock-heating mechanisms
contaminating the normal ratios expected from \ion{H}{II} regions.

The strongest equivalent width of  H$\alpha$ in the main components 
is detected in the Tail, where EW$(H\alpha) \approx 100$ \AA.  
The next highest are in the Head and Heart.
The whole measured range is typical of
\ion{H}{II} type (U)LIRGs \citep{veil95}. The strongest 
individual star formation knots falling within our slits also result in 
EW$(H\alpha) \approx 100$ \AA\ values (such as a blue double-knot in the W-wing
8 arcsec West of the Heart nucleus).
The EW values imply ages of 6 to 7 Myr for stellar populations involved,
with an instantaneous burst, and depending on metallicity \citep{monreal07}.

A rough averaged metallicity estimate can be obtained using the
N2 = [\ion{N}{II}]$\lambda$6584/H$\alpha$ ratio: from 
relations in \citet{denicolo02} we get 12+log(O/H)$= 8.85 \pm 0.12$ where
uncertainties both in the ratio and the method are included. The metallicity
would be 8.75 for the lowest detected ratio at the deblended nuclear 
component of the Heart.

The R$_{[{\rm SII]}}$ = [\ion{S}{II}]6716/[\ion{S}{II}]6731 ratio ranges 
between  
R$_{[{\rm SII}]} \approx 1.0 - 1.5$ throughout the Bird, which correspond 
to electron densities from n${_e}=$ 500 cm$^{-3}$ to less than 10 cm$^{-3}$.  
Apertures within the Heart and Body show the highest densities of 
$300 - 500$ cm$^{-3}$, while all the other Bird components 
appear to have n${_e}<20$ cm$^{-3}$ .

From the IRS mid-IR spectrum (Fig.~\ref{irs}) we measure 
a line ratio of [\ion{Ne}{III}]15.5$\umu$/[\ion{Ne}{II}]12.8$\umu$=0.10,
which is typical of low to medium excitation starbursts,
and corresponds to effective stellar temperatures of $3.5-3.7\times 10^4$ K in
\ion{H}{II} regions \citep{giveon02}. The IRS spectrum is examined
in more detail in Section~\ref{agnsfr}.

\begin{table*}
 \centering
 \begin{minipage}{160mm}
  \caption{Properties of the components of IRAS 19115-2124.}
  \label{components}
  \begin{tabular}{l l c l c c c c c} 
  \hline  
Component 
 & subcomp 
 & $R$\footnote{Approximate extent of component in the NACO image, 
   {\em not} formal $r_e$}  
 & $v$\footnote{Heliocentric line-of-sight velocity; error includes 
   formal error and the spread of velocities derived from different lines}  
 & $v_{rot}(los)$\footnote{Half of peak-to-peak velocity from rotational 
   curves of one-component fits, not corrected for inclination effects}  
 & $\sigma$\footnote{Velocity dispersion, corrected for instrumental 
   resolution and redshift} 
 & [\ion{N}{II}]/H$\alpha$\footnote{From [\ion{N}{II}]$\lambda$6583, 
    not corrected for reddening, uncertainties approximately $\pm$0.03} 
 & [\ion{S}{II}]/H$\alpha$\footnote{From [\ion{S}{II}]$\lambda\lambda$6716+6731,
    not corrected for reddening, uncertainties approximately $\pm$0.03} & EW(H$\alpha$)\footnote{Uncertainties approximately $\pm$3 \AA}\\
          &         &(kpc) &(km~s$^{-1}$) & (km~s$^{-1}$)     & (km~s$^{-1}$)   &  &  & (\AA) \\ 
  \hline
Heart...  &  total\footnote{Wide aperture, unblended multiple components}  & 4
             & 14520$\pm$20 & 185$\pm$6\footnote{The deblended adopted rotational velocity from this aperture is 157 km s$^{-1}$. See text.} &  129$\pm$7 & 0.43 & 0.35 & 51\\
          &  nuclear\footnote{Deblended component}&  0.5 & 14576$\pm$9\footnote{Adopted systemic velocity 
             of the Bird} & ... & 76$\pm$9  & 0.32 
 & 0.28 & 35 \\
          &  blue wing\footnote{Deblended blue component of the nuclear 
             aperture} & ... & 14362$\pm$23 & ... & ... & 0.56 & 0.39 & 19 \\
          &  W-disk\footnote{Separate aperture West of the nucleus}  
          & 1.5 &  14362$\pm$12 & ... & 90$\pm$6     & 0.44 & 0.41 & 37 \\
          &  E-disk\footnote{Separate aperture East of the nucleus}  
          & 1.5 &  14701$\pm$6  & ... & 52$\pm$5    & 0.42 & 0.42 & 39 \\
Body      &         &  3 &
          14590$\pm$40 
          & 75$\pm$4  & 134$\pm$16 & 0.42 & 0.32 & 42 \\
Head      &         & 2.5 &  14940$\pm$30 & 40$\pm$5 & 93$\pm$8 & 0.47 & 0.45 & 69 \\
Tail      &         & 6 &  14467$\pm$7 & ... & ... & 0.41 & 0.40 & 97 \\
E-wing    &         & 8 &  14817$\pm$6 & ... & ... & 0.45 & 0.40 & 28 \\
NW-wing   &         & 16 & 14175$\pm$6 & ... & ... & 0.44 & 0.45 & 27 \\
W-wing    &         & 3 &  14337$\pm$6 & ... & ... & 0.38 & 0.34 & 63 \\
 \hline
\end{tabular}
\end{minipage}
\end{table*}

\subsubsection{Velocities of major components}

\begin{figure*}
   \centering
   \includegraphics{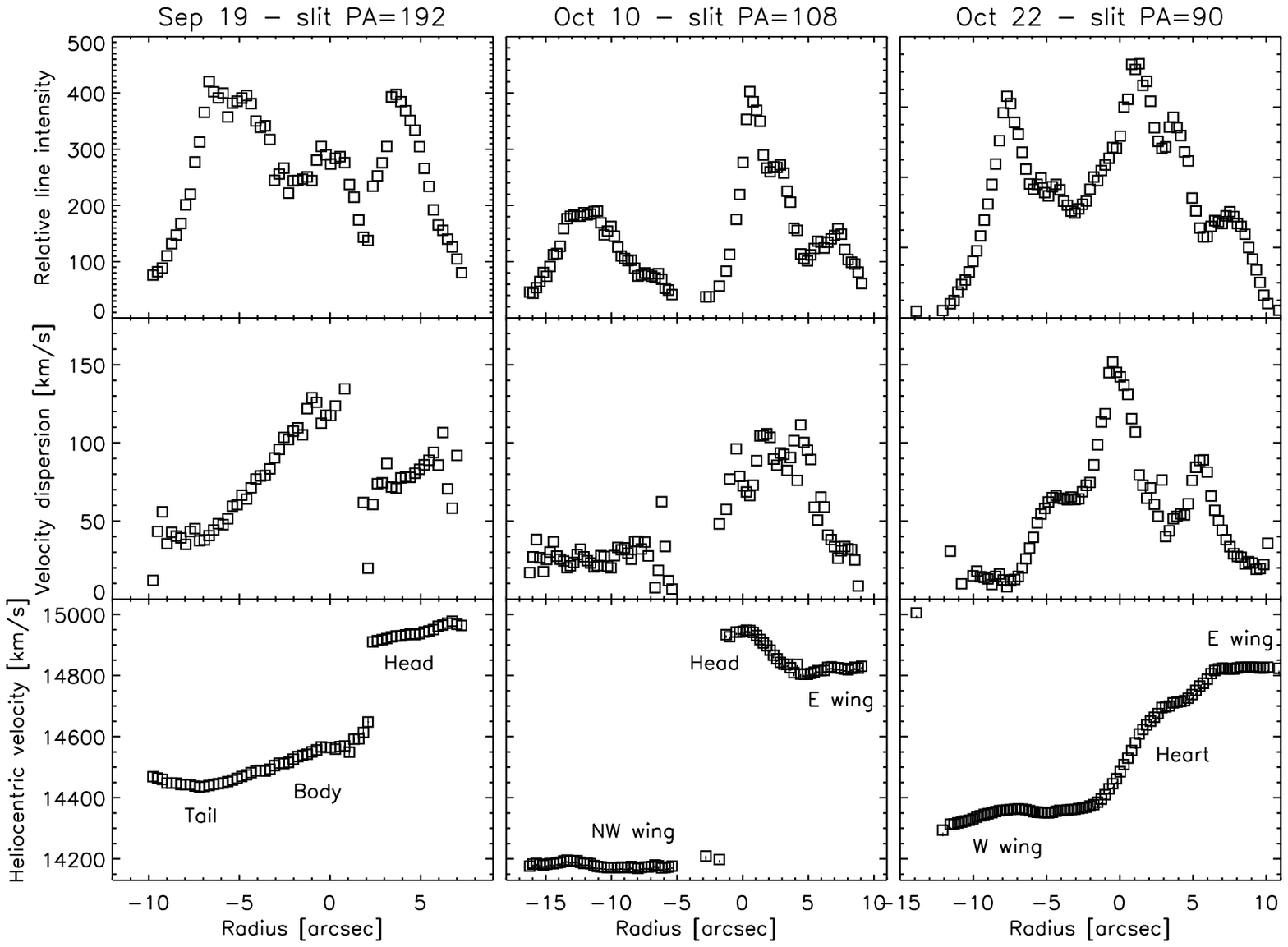}
   \caption{Rows of panels from the bottom: 
            radial velocities, velocity dispersions,
            and line intensities along the three different slits, derived
            from a single component gausssian fit to the H$\alpha$ line.}
           \label{rotdispvel}
    \end{figure*}

As seen from the RSS 2D spectra (Fig.~\ref{figha})
the velocity structure of the Bird is quite complex, and we 
used multi-component fitting to derive line ratios (above) and accurate 
velocities and dynamics (next and later sections).  
However, we found it useful to also simply fit a single-component gaussian 
to the H$\alpha$ lines along all the three slits for general 
characteristics of the primary components.
The resulting velocity curves, each point representing a single pixel row, 
are shown in the bottom panels of 
Fig.~\ref{rotdispvel}, where the main structural components are also
indicated. 
The results are unchanged, though more inaccurate, 
if the weaker [\ion{N}{II}], [\ion{S}{II}], or [\ion{O}{I}] lines are used.  

The Body and the Tail of the system join
together smoothly in velocity space, while the Head is clearly a separate
component from the rest.  The Heart, the smaller spiral, is the
only component that shows a clear rotational structure with peak-to-peak
velocity of approximately 370 km~s$^{-1}$, while any rotational velocity of the
other components is $<100$ km~s$^{-1}$ from our slit orientations.  
The velocity of the W Wing appears 
to join well the arms of the small spiral, but there is a jump of 100 
km~s$^{-1}$ from the arms on the other side to the E wing.

A general picture emerges from these spectral profiles,
where the Head is moving away from 
us with the highest radial velocity of 14 940 km~s$^{-1}$, 
i.e.\ $+360$ km~s$^{-1}$ with respect to the systemic velocity (see below),
and the E Wing has the second 
highest relative velocity of $+240$ km~s$^{-1}$. The NW Wing, in contrast, 
is turning our way with a high relative velocity of 
$-400$ km~s$^{-1}$.  The Tail has an average of $-110$ km~s$^{-1}$
offset with respect to the systemic velocity, while the Body appears
to be very close to the systemic velocity.

\begin{figure}
   \centering
   \includegraphics[width=84mm]{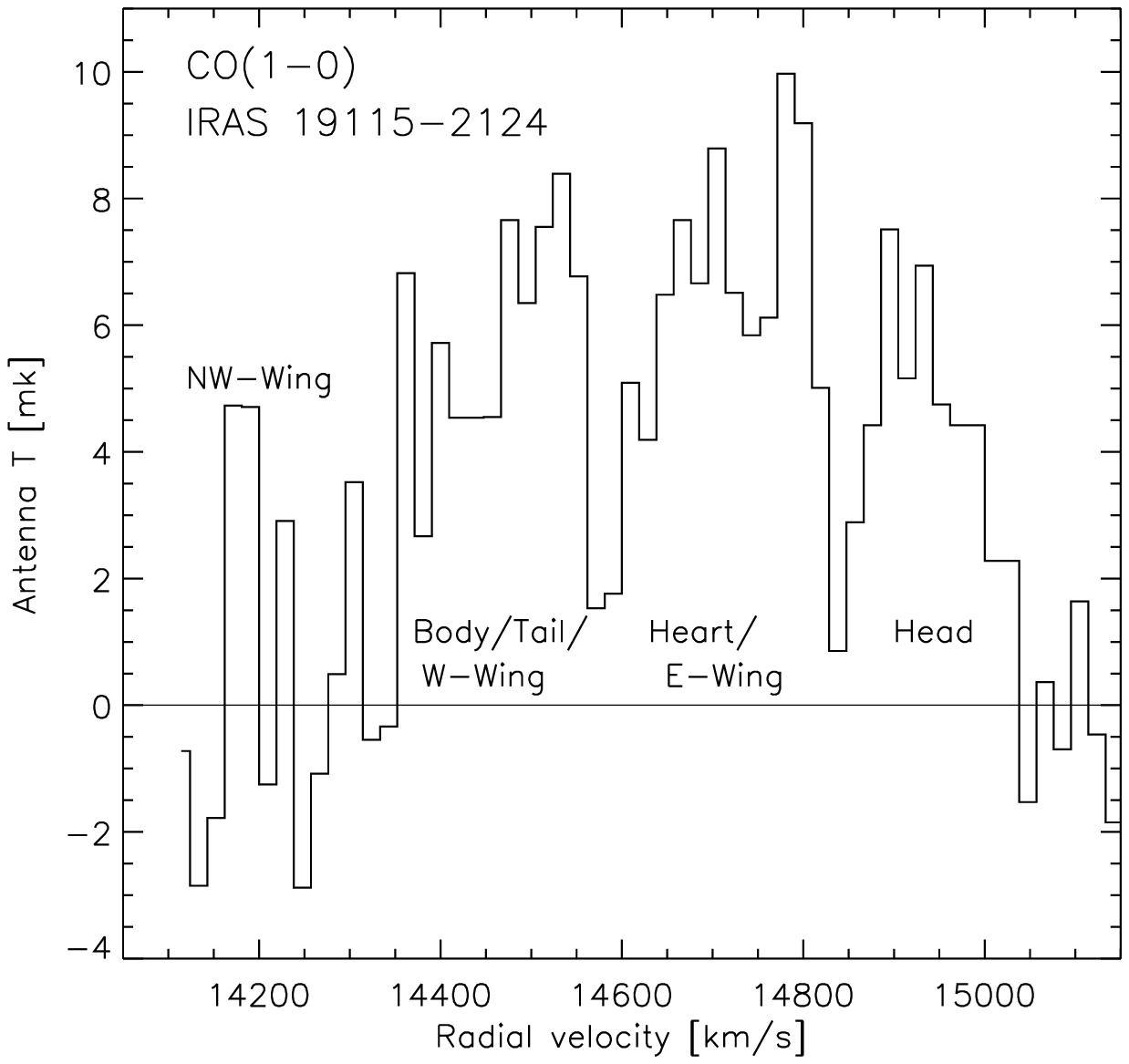}
   \caption{We have extracted the CO(1-0) observation of IRAS~19115-2124
   from of Mirabel et al.\ (1990) and labelled in the figure the velocity
   components which match the optical spectroscopy derived velocities from 
   this work. Note also the similar strenghts of CO emission, and hence
   molecular hydrogen mass, of all three components.} 
           \label{mirafig}
\end{figure}

Velocities of the different components have an excellent match with those of
\citet{mirabel90}, whose CO($1-0$) based values 
show three distinct components at 14400--14550 km~s$^{-1}$, 
14600--14800 km~s$^{-1}$, and 14850--15000 km~s$^{-1}$, corresponding to 
our Tail, Heart, and Head, respectively.
Moreover, their fourth, weaker and narrow component at 14200 km~s$^{-1}$ 
agrees exactly with our derived velocity for the NW wing of te Bird
(see Fig.~\ref{mirafig}).

We also plot the velocity dispersions (middle panels of Fig.\ref{rotdispvel})
extracted from the one-component fits along the slits.
The Wings and the Head are kinematically cold, while the Heart and Body
nuclei have quite high values (we return to velocity dispersions for 
the purpose of mass estimates in Section~\ref{masses}).
Note also that the strength of the H$\alpha$ line emission 
(top panels of Fig.~\ref{rotdispvel}) does not always
exactly correspond to the position of the maximum velocity dispersion.
The velocity dispersion peak does, however, match well 
the position of the peak 
{\em continuum} emission along the slit, which is chosen as the zero-point
of the x-axis radial distance.

Table~\ref{components} summarizes the properties of the major components
of the Bird.

\subsubsection{Velocities of sub-components and the systemic velocity}

\begin{figure}
   \centering
   \includegraphics[width=84mm]{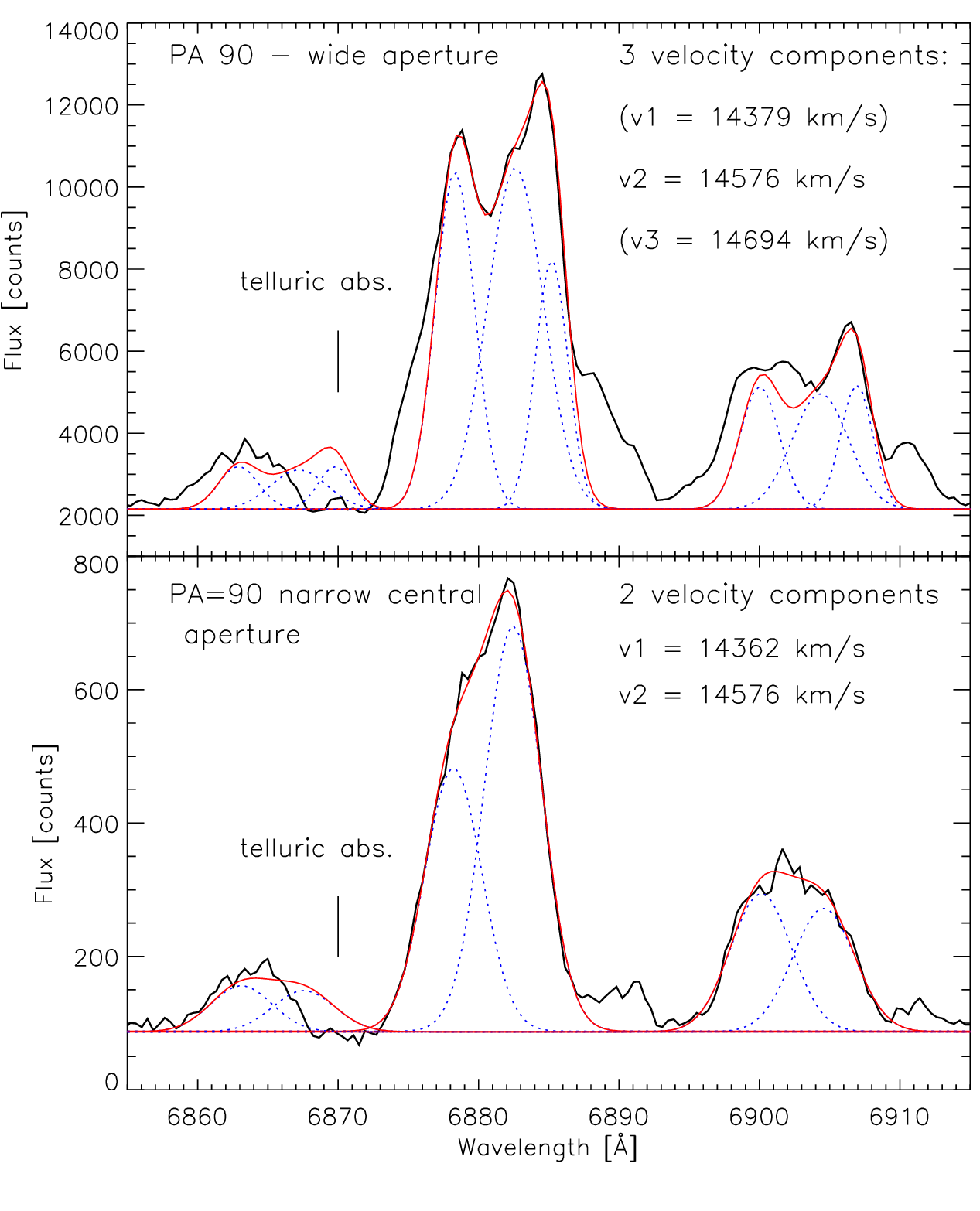}
   \caption{Extracted spectral profile (black solid line) 
   at the H$\alpha$ and \ion{N}{II} region
   of the PA=90 spectrum.  The upper panel shows the profile over a wide
   1.5 arcsec aperture at the center of the Heart galaxy, while the
   lower panel is a single pixel, 0.26 arcsec, aperture at the same
   location. Blue dotted line shows the individual
   fitted components, the red line the total fit. The  
   [\ion{N}{II}]$\lambda6548$ profile is affected by a telluric absorption 
   band at 6870\AA.}
           \label{1dspec_bothaper}
    \end{figure}

The lower panel of Figure~\ref{1dspec_bothaper} shows the spectral profile
of the October 22 data, at the wavelength region of H$\alpha$ and 
[\ion{N}{II}] lines,
over a one-pixel (0.26 arcsec) aperture at the centre of the
continuum of the observed spectrum, which we expect to correspond closely 
to the centre of the nucleus of the galaxy in the NIR image. 
A two-component gaussian is fit to the  H$\alpha$ profile, the heliocentric 
velocities of which are plotted in the figure. Exactly the same velocity 
components were fitted for the [\ion{N}{II}]$\lambda6583$ line.
The plotted [\ion{N}{II}]$\lambda6548$ curves, in contrast, are not fits, 
but are 
merely one third of the strengths of fits to the [\ion{N}{II}]$\lambda6583$ 
line, along with the appropriate wavelength shift.

The upper panel of Figure~\ref{1dspec_bothaper} shows the 1D spectrum 
profile of the same slit and same location, but summed inside a wider aperture 
of 1.5 arcsec. Wider apertures give essentially the same result. 
Comparing to the NACO image (Fig.~\ref{bigima}) it is clear that the 
aperture should include a significant part of disk and spiral arms,
in addition to the central bulge.
The main features of the  H$\alpha$ are now fitted well with 
three gaussian components.  The central component is kept fixed at the same
velocity and width as derived from the small aperture profile. A higher
velocity component not present in the small aperture profile is evident, 
and the lower velocity component has become stronger.
The same velocity components were again fitted 
for the [\ion{N}{II}]$\lambda6583$ line, 
and [\ion{N}{II}]$\lambda6548$ was treated as before.

The wider aperture profile shows a typical
double-horn structure of spiral galaxies, with an additional blended central 
component. This latter middle velocity component of $14576 \pm 9$ km~s$^{-1}$
must be from the nuclear region, since it is present identically also in 
small apertures. 
The value is close to the average of the low and high
velocity components, i.e.\ the two sides of the disk.  Independent of
these spectral fits, we also 
derive a value of approximately 14560 km~s$^{-1}$ for the point of symmetry 
in the rotation curve.  Since the Heart of the Bird is the central component
of the whole system, and since this velocity corresponds spatially to the
$K$-band continuum peak, we adopt this  $14576 \pm 9$ km~s$^{-1}$
as the systemic velocity of IRAS~19115-2124.

Note that the strongest H$\alpha$ flux does not coincide spatially 
with the location of systemic velocity, i.e.\ the continuum peak position 
(see right-most panels of Fig.~\ref{rotdispvel}). We measure
$v = 14622 \pm 9$ km~s$^{-1}$ for the deblended velocity at the strongest 
H$\alpha$ location, approximately 
1.0 arcsec East of the nucleus.  Without the 
high-resolution AO image, this would have naturally been interpreted as the
central velocity. Indeed, the literature systemic velocity of 
$14608 \pm 48$ km~s$^{-1}$ \citep[NED,][]{strauss92} agrees well with this 
strongest H$\alpha$ component. 
It should be noted, however, that 
none of these values agree with the NED values for individual 
IRAS~19115-2124 N and S components from \citet{kim95}. 

Returning to Fig.~\ref{1dspec_bothaper}, upper panel,
there are some additional velocity components we did not fit
in the Heart central regions.  In the wide aperture extraction 
there are two weaker structures at 14820
and 14950 km~s$^{-1}$ which are contamination from the Head and the E-Wing.
But more significantly, the blue side of the lowest velocity (v1)  
component is not well fitted: the FWHM=3.1\AA\ of this component is wider 
than the 2.7\AA\ of the higher velocity arm, already implying contamination,
and the fit still leaves excess emission.
Moreover, the blue excess wing is present also in the small 
aperture profile (lower panel), i.e.\ in the nucleus only aperture. 
We interpret these
effects as outflowing gas present in most apertures over the system, 
blue-shifted here by velocities exceeding 150 km~s$^{-1}$. 
That the blue component has an elevated [\ion{N}{II}]$\lambda$6383/H$\alpha$ 
ratio may be a result of contribution from shock-heating 
\citep[][]{veilrupke02,monreal06}.

\subsubsection{Outflows detected in Na D absorption}
\label{outflows}

\begin{figure}
   \centering
   \includegraphics[width=84mm]{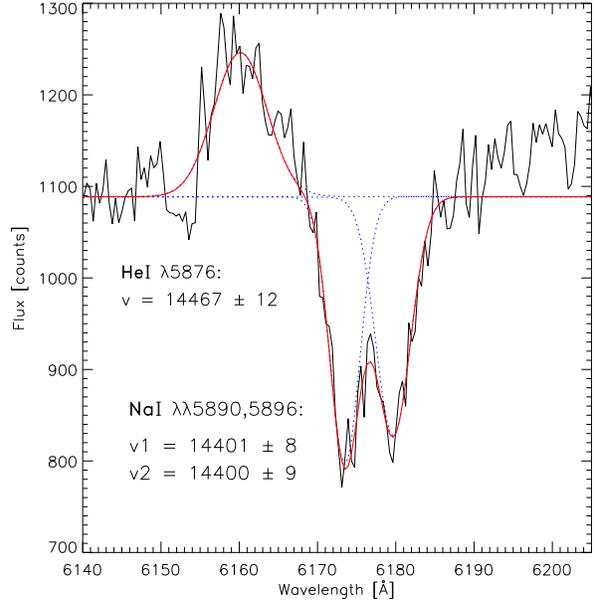}
   \caption{A three-component fit (red solid line is the total
            of dotted blue individual fits) to the 
            \ion{Na}{I} D $\lambda\lambda$5890,5896 absorption doublet 
            and the \ion{He}{I} emission line in the
            PA=90 slit at the nuclear region of the Heart.
            The derived velocity of the absorbing component, 
            14400 km~s$^{-1}$, 
            fits well the excess blueshifted emission in H$\alpha$ and 
            [\ion{N}{II}] lines (Fig.~\ref{1dspec_bothaper}).}
           \label{sodium-line-1}
    \end{figure}

Additional evidence for the outflows come from strong \ion{Na}{I} D 
$\lambda\lambda5890,5896$ (NaD) absorption
doublets in the spectra, the strongest case of which is seen in the 
central aperture of the Heart shown in Fig.~\ref{sodium-line-1}. 
In strongly star-forming systems, the NaD absorption originates partially,
or mostly, from ISM, rather than being only of 
stellar origin, and it has often been used specifically to study outflows from 
star-forming galaxies and (U)LIRGs 
\citep{phil93,sparks97,heck00,rupke02,schw04,martin05},
and to isolate kinematics of different gaseous and stellar components 
within the galaxies \citep[e.g.][]{arribas07,brosch07}.

The measured velocity of 14400 km~s$^{-1}$ for the doublet coincides 
with the blue
wing of the nuclear aperture profile of the H$\alpha$ and [\ion{N}{II}] lines,
implying an outflow velocity of the gas of approximately 180 km~s$^{-1}$ 
at this location.

Most slit positions and apertures show NaD absorption, usually
blueshifted from the emission line positions.  
The doublet profile is not always as clear as in the case of the Heart, 
but rather shows a wide range of velocities, and/or the presence of
a wider absorption component presumably originating from zero-velocity
stellar component.  Table~\ref{outflowtable} 
summarizes the outflow characteristics in the various components of the Bird,
based on NaD.  Where possible, we have also extracted the equivalent width 
ratio ${\rm R_{Na}}$ = EW(5890) / EW(5896), which is a measure of optical depth
\citep{rupke02, schw04}: ${\rm R_{Na}}=2$ for optically thin medium,
while the ratio approaches unity with infinite optical depth.

Overall, the strongest outflows are related to the 
locations of the Heart and Head, and they are somewhat weaker over the Body.  
If the outflows were from the major galaxy nuclei, the Heart and Body, 
the implied radial velocities are in the range 100--300 km~s$^{-1}$.  However,
the velocities can be up to 600 km~s$^{-1}$ if some of the outflowing material
originates from the Head.  Any outflows at the positions of the Wings or the 
Tail are below our detection limits.

\begin{table}
 \centering
 \begin{minipage}{84mm}
  \caption{Outflows and related velocities from IRAS 19115-2124 components.}
 \label{outflowtable}
  \begin{tabular}{l c c l}
  \hline  
Component &  NaD doublet\footnote{Offset velocity of the NaD doublet 
             relative to the component's velocity from 
             emission lines. '0' means
             a zero-velocity feature, and '--' a non-detection.} 
          &  NaD range\footnote{Velocity range of a wider NaD absorption 
             feature.}
          &  Optical depth\footnote{From the ${\rm R_{Na}}$ 
             = EW(5890)/EW(5896) ratio.
             '--' means the ratio can not be measured reliably.}\\
          &  (km~s$^{-1}$)     & (km~s$^{-1}$) &   \\
  \hline
Heart     &  180  &  $180-400$\footnote{In a wider aperture than shown 
                                      in Fig.~\ref{sodium-line-1}} & thick \\
Body      &  --   &  $0-270$  &  thick \\
Head      &  300  &  $0-600$  &  --    \\
Tail      &    0  &  --       &  --    \\
E-Wing    &    0  &  0        &  thin  \\
NW-Wing   &    0  &  0        &  --    \\
W-Wing    &   --  &  --       &  --    \\
 \hline
\end{tabular}
\end{minipage}
\end{table}

\subsection{Extinction estimate and gas and dust content}

Since our RSS spectra did not reach the H$\beta$ line, we first estimate an
average extinction from the emission line strengths of
IRAS~19115-2124 from \citet{kim95}: 
their H$\alpha$/H$\beta \approx 9.0$\footnote[3]{The value is from 
their North component of the galaxy, since
their South component clearly suffers from a simple line misidentification.},
implying A$_V \approx3.3$, or A$_K \approx0.36$, which is typical of 
local ULIRGs. Of course, it is obvious 
from the optical images that extinction is highly variable over the Bird,
for example the Body and southern parts of the Head are most
affected by dust, but as an averaged value the above indicates significant, 
yet not extreme extinction.  This value is consistent with a
statistical value found from an empirical relation found for IR-luminous 
galaxies in the {\em Spitzer} First Look Survey \citep{choi06}:
A$_V = 0.75 \times \log(L_{IR}/L_{\odot}) - 6.35$ which would give 
A$_V = 2.5$ for the Bird.

We also make an extinction estimate based on the detected 
NaD absorption feature -- 
this can be done assuming it is mostly of interstellar origin, rather than 
stellar. \citet{veil95} show that there is a correlation 
between EW(\ion{Na}{I}D) and colour excess E(B-V) for \ion{H}{II} type 
(U)LIRGs \citep[see also][]{sparks97}.  Taking EW(\ion{Na}{I}D) $\sim 
5\times$ E(B-V) and our EW(\ion{Na}{I}D)$\approx4$~\AA\ from the strongest 
case in the Heart, gives E(B-V)~$\sim$~0.8, and hence A$_V \sim 2.4$.

Broad-band colours give indications of dust reddening in specific
regions. Using photometry in 500~pc radii for the Body and Heart nuclei 
(Table~\ref{phottable}) and in 1 kpc radius for the brightest NIR feature 
in the Head, we measure $I-K = 4.4$, 3.8, and 3.7, 
respectively, and $B-K = 8.4$, 7.1, and 6.6. 
Since normal spiral galaxy colours are approximately $B-K\sim$3.5 and 
$I-K\sim$2, these colours imply A$_V\sim$4--5, i.e. somewhat higher 
extinctions than from the methods above. 
Both the Body and Heart regions would in fact nominally be classified as
Extremely Red Objects, EROs \citep{petri}.

The {\em Spitzer} data has the potential of probing the most extincted regions.
\citet{leven07} have calculated models 
that relate optical depths to IRS spectral 
features:  the strength of the MIR silicate feature 
(see Fig.~\ref{irs}) defined as 
$S_{\rm SIL} = \ln{ [f_{obs}(9.7\umu m) / f_{cont}(9.7 \umu m)] }$,
where $f_{obs}$ and $f_{cont}$ are the observed flux density and the
estimated continuum flux density,
$S_{\rm SIL} \approx -1.4$, implies optical depths
in the range $\tau_{V} = 30 - 200$ depending
on the dust shell thickness and geometry. This range can be constrained by  
making use of the rest wavelength continuum
ratio, $f_{\nu}(14\umu{\rm m}) / f_{\nu}(30\umu{\rm m}) \approx 0.15$; 
the geometric models giving consistent
optical depths with both indicators simultaneously are ones with very thick
$R_{out} / R_{in} > 200$ and nearly uniform radial density distribution  
($p\sim0$ for $r^{-p}$), which indicate 
extinction at the lower end of the range, $\tau_{V} \sim 30$ (A$_V\sim 30$).
Thus, the MIR spectrum extinction 
indicators show the Bird to be a moderately extincted source 
as far as (U)LIRG nuclei are concerned.  

Finally, we make an estimate of the gas masses involved.
Taking an optical extinction estimate of A$_V = 4$ and using 
a standard gas-to-dust relation of N(H)/A$_V \sim 1.9 \times 10^{21}$ atoms 
cm$^{-2}$ mag$^{-1}$ \citep{bohlin}
we get n(H) = $1.3 \times 10^{-5}$ kg cm$^{-2}$, so
assuming an absorption layer covering the whole system, say, a 3 by 8 kpc 
area, the gas mass would be $M \sim 1.5 \times 10^9 \ {\rm M}_{\odot}$.   
If, instead, the higher optical depth of A$_V = 30$ were used in 
conjunction with a 2 kpc 
diameter area of the 8 and 24 $\umu$m emission around the Head and Heart 
components, we would arrive at $M \sim 2 \times 10^9 \ {\rm M}_{\odot}$.
These are reasonable values recalling that the CO measurement of molecular
hydrogen by \citet{mirabel90}, probing the full emitting volume regardless 
of geometry and optical depth, gave $3\times10^{10} \ {\rm M}_{\odot}$.

\subsection{Masses and dynamics}
\label{masses}

\subsubsection{NIR light based mass}

Before estimating the dynamical mass of IRAS 19115-2124 and its components, 
we use the absolute K-band magnitudes presented above in 
Section~\ref{photometry}.  The total integrated absolute 
magnitude from our NACO image is $M_K = -25.5$ mag,
or about $3.6\times L^{\star}$.  The individual brightnesses in 1 kpc
aperture radii for the
Body, Heart, and Head are $M_K=-23.7$, $-23.2$, and $-22.7$ mag, respectively.
These translate to 6.3, 4.0 and $2.5\times 10^{10} {\rm L}_{K,\odot}$
(or 0.7, 0.4, and 0.3 $L^{\star}$, and nearly 2/3 of the total Bird
brightness is thus contributed from outside the 1 kpc apertures).
It is commonly assumed that NIR light traces reasonably well stellar mass 
(though the quantitative relation is wrought with complications) -- as an
estimate, we use a $K$-band mass-to-light ratio from \citet{thron88}, and
calculate $3.0\times 10^{11} {\rm M}_{\odot}$ for the whole Bird system 
($K$=11.0 mag, and D$_L$=207 Mpc).
Taking $m_{\star} = 1.4\times 10^{11} {\rm M}_{\odot}$, 
this corresponds to a mass of 2.1~$m_{\star}$ in (old) stellar
population in the Bird. The three individual components, within 1 kpc radii
each, would range from 0.16~$m_{\star}$ (Head) to 0.40~$m_{\star}$ (Body),
see Table~\ref{dynpar}.  Since the component
masses were calculated within 1~kpc radii the likely total 
masses would perhaps be a factor 2 larger, in accordance with the total-light 
derived value. 

No extinction corrections were made above, making these estimates lower
limits. Average extinctions over the major nuclei are in the range 
A$_V \sim 2.5-5$, as seen above, which would mean upward mass corrections
of 30 to 60 percent.  However, we have not attempted a correction due to  
light from bright, recently-formed red supergiant populations, the continuum 
light of which would in contrast result in us predicting too much mass: 
their contributions 
range from nearly negligible in normal spiral galaxies to perhaps 40 per cent 
in localised young starburst regions \citep[e.g.][]{rhoads,james}. The 
correction may be significant in case of the Head, where 
intense individual (presumably young) clusters are observed in the NIR, but
it is unlikely to be very significant in the Heart and Body components.  
As a further potential correction, based on typical NIR spectra of ULIRGs 
\citep{murphy99}, we estimate the expected contribution from emission lines 
(especially Pa$\alpha$, Br$\gamma$, and H$_2$ vibrational series) to the 
broad-band light to be {\em maximally} $\sim10$ per cent, but likely much 
less than 5 per cent, since the strongest contamination from Pa$\alpha$
only marginally enters the edge of the NACO $K_s$ filter at the Bird redshift. 
Taken together, it is more likely that an upward correction to mass 
due to extinction dominates over any downward 
corrections due to starbursts and emission lines from gas.  
Since all the corrections are somewhat uncertain, but nevertheless of similar
magnitude, we prefer to 
keep the $K$-band light mass estimate as is, as a lower 
limit, without applying the corrections.  
Finally, note that the mass-to-light conversion used implies values 
of M/L$\sim0.9$ in the system: this is consistent with values
expected from stellar populations in high-luminosity (isolated) spirals 
calculated from models that also include moderate starburst episodes 
\citep{bell01}.

\subsubsection{Dynamical mass}

For dynamical mass estimates we follow \citet{dasyra06b}
\citep[see also][]{colina05}  
and express $m$ (in units of $M_{\odot}$) as
\begin{equation}
  m = 4.7 \times 10^5 \ (3\sigma^2 + v_{rot}^2) \ r_e
\label{mass}
\end{equation}
where $\sigma$ and $v_{rot}$ are the line-of-sight velocity dispersion and the
inclination corrected rotational velocity, both in units of km~s$^{-1}$,
and $r_e$ is the effective radius in kpc.  The relation assumes a virialised
central region, and the constant term depends on the distribution
of matter in the system: the value above assumes a King model with
constant mass-to-light ratio, with tidal-to-core radius ratio of 50, which
is midway between values for dwarf and giant ellipticals.  Note also that
the rotational velocity component effectively adds the mass of any disk to 
that of the bulge component estimated by the velocity dispersion term.

To get the true rotational velocity $v_{rot}$ one has to correct it both for 
deviations of the slit orientation from the major rotational axis 
and inclination effects.  In the following $v_{rot}(los)$
is the line-of-sight rotational velocity from our data, $\theta$ is the
angle between the slit and the major axis of rotation and $i$ the inclination.
\[ v_{rot}  =  v_{rot}(los) / (\cos(\theta) \sin(i)).  \]

The only case of significant rotation seen in the three slit
positions, is the Heart.  The best-quality slit position PA=90 
accross the Heart is aligned nearly perpendicular against the major rotation
axis, in principle making the required orientation correction 
difficult.  However, the 1.5 arcsec slit-width actually encompasses 
a large fraction of the disk area, which we estimate from the NACO image as
approximately 2.0 arcsec across, and therefore we can estimate which fraction
of the maximum rotational velocity enters the slit.  Based on this geometrical
assessment, we adopt an approximate value of $\theta \approx 30$ in the above 
equation.  We adopt $v_{rot}(los) = 157$ km~s$^{-1}$ 
measured from the fitted blue 
and red velocity components in the central 1.5 by 1.5~kpc region of the 
Heart (Fig.~\ref{1dspec_bothaper}.
We thus have $v_{rot}(obs) = 181$ km~s$^{-1}$ as the observed rotational 
velocity. 
For an inclination estimate we use the elongation of the Heart disk,
which has the ratio of the minor-to-major axes $b/a \approx 0.75$ from NACO
isophotes, i.e. $i \approx 44 \deg$, assuming 0.3 for the ratio of thickness 
and truncation radius of the disk.  Thus, $v_{rot} = 261$ km~s$^{-1}$, 
which is a 
reasonable value for fairly massive disks.  The rotational velocity would
be $v_{rot} = 226$ km~s$^{-1}$ if only the inclination correction is applied.

Similarly, we estimate $v_{rot} = 75$ km~s$^{-1}$ for the Body, 
and $v_{rot} = 40$ km~s$^{-1}$
for the Head, both from the PA=192 data.  For the Body $\theta \approx 0$,
and inclination $i\approx60$ were used, whereas no corrections to the apparent 
observed $v_{rot}(los)$ of the Head were made due to irregularity and
ambiguity of its shape.

We next measure velocity dispersions in 1.0--1.5 kpc apertures at the centers 
of the galaxies, corresponding to their adopted systemic velocities, with 
multiple component Gaussian fits to H$\alpha$, 
[\ion{N}{II}]$\lambda$6583 and [\ion{S}{II}] lines, and quadratically correct 
them for our instrumental resolution, and adjust for redshift 
(Table~\ref{dynpar}). Single component fit results, which may be 
contaminated by gas outflows, are also tabulated.
The Heart and Body velocity dispersions come out to be $\sigma = 76 \pm 9$ 
km~s$^{-1}$ and $\sigma = 94 \pm 10$ km~s$^{-1}$, respectively, 
which can be considered as definite lower limits for velocity dispersion.

Now, with the values of rotational velocities and velocity dispersions at 
hand, we may estimate the masses using Eq.~\ref{mass} (see Table~\ref{dynpar}).
For the sizes $r_e$ we trust the
fits over 1D cuts along major-axes more than the
GALFIT best-fitting values, which are very sensitive to e.g.\ choices of 
sky background and exact Sersic $n$ parameter. 
Half-light radii measured photometrically from the NACO image are typically
close to the lower range of $r_e$ values of both the 1D cuts and GALFIT 
for the Body and Heart. 
We conservatively estimate the uncertainties to be at about 
30 per cent level.  In the case of the Head, the irregular 
morphology prohibits a formal $r_e$ fit, and the value tabulated is an
estimate based on half-light radius from photometry. 
  
The dynamical mass of the Heart, using Eq.~\ref{mass}, comes out to be 
0.36~$m_{\star}$ (0.50~$m_{\star}$) using the deblended (unblended) 
velocity dispersion.
Since this is a clear case of a spiral galaxy, 
we also tabulate the mass obtained directly from a simple Newtonian system 
enclosing a mass inside 3 kpc radius and rotating at 261 km~s$^{-1}$.  
This estimate gives a lower limit of 0.32~$m_{\star}$ for the Heart system, 
with the caveat that some 
part of the adopted $v_{rot}$ may be tidal rather than orbital motion.  
The dynamical mass estimate of the Body is 0.27~$m_{\star}$ 
(0.50~$m_{\star}$).  This system is dynamically
hot -- ignoring the rotational component does not appreciably change the 
mass estimate.  Finally, the corresponding dynamical mass for the Head 
is 0.08~$m_{\star}$.

The mass ratio of the three main components of the Bird thus becomes
4.5:3:1 (6:6:1) for Heart:Body:Head when the deblended (unblended) velocity 
dispersions are used.  If the rotational 
velocity estimate only is used for the Heart, the ratio is 4:3:1, and
when $K$-band light based estimate is used, the ratio is 1.5:2.5:1.
This last ratio is the only one measured within identical radii.
Every option nevertheless implies two close to equal major merger components, 
and a third component in the range 20 to 40 per cent in mass of the 
more massive ones.

\begin{table*}
 \centering
 \begin{minipage}{140mm}
  \caption{Dynamical parameters and mass estimates for IRAS 19115-2124. 
           We adopt $m_{\star} =1.4 \times 10^{11} {\rm M}_{\odot}$.}

 \label{dynpar}
  \begin{tabular}{l l c c c c c c c} 
  \hline  
Component &  note & $M_K$  & $r_e$\footnote{We estimate the systematic uncertainty to be $\approx30$ per cent in the adopted effective radii.}  & $v_{rot}$  & $\sigma$ & $v_{rot}/\sigma$ &$m_{light}$\footnote{NIR light measured in 1~kpc radius, see text for M/L ratio, which also dominates uncertainty.} &  $m_{dyn}$\footnote{Uncertainty is $\approx40$ per cent, 
dominated by the uncertainty in $r_e$} \\
          &       & (mag) & (kpc)  & (km~s$^{-1}$) & (km~s$^{-1}$) &  &$(m_{\star})$ & $(m_{\star})$ \\
  \hline
Heart       & \footnote{Deblended velocity dispersion} &   ``     &  ``          & ``   &  76$\pm$9 & 3.4 & 0.25 & 0.36  \\
...         & \footnote{Single component velocity dispersion} &  -23.16$\pm$0.07  &  1.26  &  261$\pm$35 & 129$\pm$7 & 2.0 & 0.25 & 0.50 \\
...     & \footnote{Using Keplerian rotation at 3 kpc}       &   ``     &  3.0          & 261$\pm$35  &  -- & --  & 0.25 & 0.32  \\
Body     & {\em d} &   ``     &  ``          & ``   & 94$\pm$10 & 0.8 & 0.40 & 0.27  \\
        & {\em e} &  -23.69$\pm$0.07  &  2.52  &   75$\pm$7 & 134$\pm$16 & 0.6 & 0.40 & 0.50  \\
Head    &  &  -22.71$\pm$0.07  &  0.84       &   40$\pm$5 &  93$\pm$8 & 0.4 & 0.16 & 0.08        \\ 
 \hline
\end{tabular}
\end{minipage}
\end{table*}

\section{Discussion}

\subsection{Dynamics and kinematics}

\subsubsection{Reliability}

Can masses determined from velocity dispersions 
and rotational velocities calculated from optical emission lines 
be trusted for dynamical 
mass estimates?  The answer depends very much on how closely the velocities
and structures of warm ionized gas
are coupled to the stellar stuctures and velocities.
From the outset, there is no reason to expect closely matching correlations
since the optical emission lines trace regions well outside the more
extincted nuclei and since the very clumpy and disturbed nature 
of nuclear regions in (U)LIRGs make correspondences highly complicated.

Recently \citet{colina05} presented a detailed study 
of a sample of 11 ULIRGs with angular resolutions of approximately 1 kpc 
throughout, and covering a range of ULIRG types,
utilizing optical integral field spectroscopy (tracing the gas distribution 
with H$\alpha$ light), medium-resolution Keck and VLT near-IR spectroscopy 
(tracing stellar light with CO absorption bands), and millimetre CO 
observations (tracing the cold gas structures).  They find, as expected, that
in general over a few kpc the ionized gas has a complex velocity structure, 
with the peak-to-peak
velocity differences, and often also velocity dispersions, dominated by tidal 
tails and tidally induced flows, rather than rotational systems 
(though they do find ionized gas matching stellar
velocities much more accurately than the molecular gas). However, 
they also show that the {\em central} 
warm gas velocity dispersion correlates well with the 
stellar velocity dispersion (they find a ratio of $1.01 \pm 0.13$) and
conclude that the central ionized gas velocity dispersion is, in fact, 
a robust tracer of dynamical mass of these systems, {\em provided} the
emission lines in question can be spatially associated with the 
true nuclei identifiable
virtually only by high-resolution near-IR imaging. We also note that
\citet{rupke02} found a good correlation between the widths of emission lines,
the Ca II triplet widths, and of the absolute K-band brightness of their
sample of ULIRGs. 

During this study we have seen exactly the problems posed in the 
\citet{colina05} work -- without the extremely high quality near-IR
imaging it would have been impossible to determine which features in the
optical spectra correspond to nuclear 
and surrounding locations. For example, we are now able to tie the systemic 
velocity to the nucleus of a major component, rather than a bright off-centre
\ion{H}{II} region where the literature value seems to have originated from.
We have furthermore been very conservative in 
deblending velocity structures in spectral profiles within the NIR 
AO-determined nuclear regions to account for outflows and thus 
avoided over-estimating masses by factors of 2--4 because of them.  

We thus conclude that the optical 
spectroscopic method of defining kinematics and dynamics of the Bird system, 
in conjunction with NIR AO imaging, should be robust and reflect true dynamical
masses within observational errors.

\subsubsection{Tully-Fisher relation}

The Heart spiral galaxy has an absolute brightness of $M_K = -23.2$ and 
its rotation velocity is $V_{max} = 260$ km s$^{-1}$.  Placing these values
into a $K$-band Tully-Fisher relation diagram \citep[][]{conse05}, 
we see that it lies well inside the expected 3$\sigma$ region of normal disk 
galaxies, on the fainter and/or faster rotation side of the
relation;  the small offset might be partially
explained by under-estimated brightness due to extinction, or perhaps
by dynamical effects due to the on-going interaction.

\subsubsection{The Fundamental plane}

Averaged within the respective half-light radii of the Heart (1.2 kpc) and 
Body (2.5 kpc) components, we calculate $K$-band surface brightnesses of 
14.8 mag arcsec$^{-2}$ and 15.8 mag arcsec$^{-2}$.
Positioning these on the Fundamental plane of elliptical galaxies, we find 
them both to occupy typical regions of (U)LIRGs \citep{genzel01,dasyra06b}, 
i.e.\ well within the scatter of the relation over different types of objects, 
but typically 1--2 mag brighter than
ordinary hot E/S0 galaxies.  That our nuclei,
together with other (U)LIRGs, occupy regions relatively close to
the Fundamental plane is commonly interpreted 
as evidence that these objects are ellipticals in formation.

\subsubsection{Escape velocities}

Following \citet{rupke02} we define an escape velocity $v_{esc}$ at
radius $r$ for a singular
isothermal sphere truncated at $r_{max}$ as
\[ v_{esc} = \sqrt{2} v_c [1 + \ln{(r_{max}/r)} ]^{1/2} \]
where $v_c$ is the rotation speed of an object in orbit.  The escape
velocity is not very sensitive to the choice of $r_{max}/r$, though 
this ratio is likely
to be large ($>10$) since the dark matter halo extends far from the 
optically observable galaxy. Choosing  $r_{max}/r = 10$ and using
$v_c \approx 260$ km~s$^{-1}$ (Heart rotation) as the extreme case, 
we get $v_{esc} \approx 670$ km~s$^{-1}$.  If the radial velocities of the
components are close to the true relative velocities, it would be 
unlikely that any significant fraction of the outflow material or sub-systems
are escaping into the intergalactic medium.
The 350--400 km~s$^{-1}$ relative radial velocity offsets of the Head and the 
NW Wing are, however, sufficiently close to the $v_{esc}$ that it
is difficult to say whether they are truly bound or not.
If one considered only the velocity dispersions
in the two nuclei, which are at most $\approx 130$ km~s$^{-1}$, 
the escape velocity
would be $v_{esc}=330$ km~s$^{-1}$ meaning that the Head, Wings, and
much of the outflows would be heading out of the system based on radial 
velocities only.

\subsection{Classifying the Bird sub-systems}

\subsubsection{Differentiating nuclei from star clusters}

The Bird consists of two massive components, one still a clear
spiral, and the other a disturbed disk/bulge system, 
of two tidal tails/wings, and an 'extra', but rather massive, 
irregular component with many individual bright star-forming regions.

Though there have been reports of a significant fraction of (U)LIRGS being
mergers of multiple components \citep{borne00,cui01}, most surveys  
tend to agree that they usually are interactions and mergers of two
galaxies, with the incidence of multiple mergers less than 5 per cent 
\citep{veil02,bus02}.  The complicating factor, acknowledged in all 
(U)LIRG studies, is that any studies made in the optical regime, 
even at I-band, suffer from very complex extinction effects making photometric
or morphological identification of nuclei difficult.
In the case of the Bird, if only the $I$-band were considered, 
we would have 6 bright clumps (measured in 600 pc diameter radii) 
of $M_I<-17$ mag, which was selected as the cut-off between
legitimate nuclei and giant star-forming regions in \citet[e.g.][]{cui01}. 
Only our NACO image
reveals that three of those clumps are in fact the nucleus and two spiral
arms of the same galaxy, and one of the clumps is a bright knot within the tail
of the Bird, clearly a minor feature in the system.

\subsubsection{The nature of the Head of the Bird}

It is the more irregular feature, the Head of the Bird, which we are now 
concerned with. Is it a legitimate nucleus, a remnant of a pre-existing 
galaxy, or merely a collection of star clusters, formed out of
the material of the other two main components during the interaction, 
for example a tidal dwarf galaxy (TDG)?

It is clear that the differences in brightness are not that large between the
three components.  In the B-band and I-band, the Head is in fact the brightest
single feature in the whole interacting system. Even in the 
$K$-band, which more closely follows the underlying mass, 
the Head is only 1 mag fainter than the
brightest component of the whole system.
The dynamically derived mass for the Head ranges from one-third to one-sixth
of the most massive component, depending on the method used.  
The CO-line measurements of \citet{mirabel90}
indicate an H$_2$ mass for the Head of not less 
than half of the most massive H$_2$ component (Fig.~\ref{mirafig}).  
Finally, even the brighest NIR knot alone, which is totally
obscured in the $B$-band image, 
has $M_K \approx -20.4$ and $M_I \approx -16.7$, 
making it very close to the e.g.\ the \citet{cui01} 
criterion for individual nuclei.

Furthermore, the Head is kinematically 
totally separate from the Heart and Body structures,
with a difference in velocity nearly 400 km~s$^{-1}$ 
(see Fig.~\ref{rotdispvel}).  It also has evidence of rotation,
suggesting a self-gravitating body.  
The appearance of the optical morphology is
that of two gas rich disks colliding almost perpendicularly, with the 
NIR image revealing the Heart nucleus to be likely associated with the 
horizontal disk component (the Wings) and the Body nucleus being more likely
associated with the vertical structure comprising also the Tail
-- and potentially the Head.  However, given the clearly discrepant velocity
(and the significant stellar mass of the Head), it is difficult to imagine the 
Head being material merely ripped off a galaxy that originally had the Body
component as its nucleus, and that component remaining such a conspicuous
single nucleus.

In fact, the optical images show a large obscuring dust region crossing 
perpendicularly the E-Wing, and also extending some
6 kpc to the E and NE of the Head. This structure looks as if
it could be separate from the two major horizontal and vertical components;
and if this is true, it could be material from the same structure as the 
stellar component associated with the Head. The gas velocity 
($\sim15000$ km~s$^{-1}$)
of this structure can be traced 4 kpc North out of the main Head area in NIR
(PA=192; Fig~\ref{rotdispvel}) and together with the velocity at the 
eastern tip of the E-Wing ($\sim14830$ km~s$^{-1}$), 
suggest a common kinematic origin with the Head ($\sim14950$ km~s$^{-1}$).

Though the final answer warrants detailed 
dynamical merger simulations out of the scope of this paper, we propose that
the data presented here strongly suggests a near-ULIRG merger of three 
pre-existing major components.  The Bird would be, to our knowledge, the
most-detailed studied case thus far.  For a possible triple merger
case of a lower luminosty LIRG see \citet{lipari00}.

\subsection{AGN contribution and star formation rate}
\label{agnsfr}

We discuss next whether there is a possibility of an AGN contribution
to the overall luminosity of the Bird, based on both optical and
MIR considerations.  First of all, there are no broad lines anywhere 
in the optical
RSS spectra, nor do the line ratios suggest clear AGN activity.  Some 
line ratios close to the \ion{H}{II} vs.\ LINER dividing region come from areas
associated with outflows, and thus may be attributed to shock-heating. 
On the other hand, while most IRAC mid-IR colours of the Bird are those of 
normal spiral galaxies, along with strong PAH emission expected in starbursts
contributing to the 8 $\umu$m band, the Head is 
redder than the other components in all IRAC colours, suggesting a hotter dust 
contribution.  

To differentiate between AGN and starburst origin, we examine
the IRS spectrum. Following \citet{sturm02, verma03, farrah07} we calculate
diagnostic line ratios: [\ion{Ne}{III}]/[\ion{Ne}{II}]~$= 0.11 \pm 0.03$, 
[\ion{S}{IV}]/[\ion{S}{III}]~$= 0.6 \pm 0.2$ (the [\ion{S}{IV}] line is totally
within the wide silicate absorption feature), 
and [\ion{O}{IV}]/[\ion{Ne}{II}]~$\approx 0.03 \pm 0.02$
(the [\ion{O}{IV}] line is a rather marginal detection). 
The Neon line ratio by itself, and also together with the Sulphur line ratio,
indicates a medium excitation starburst.  The marginal detection of 
[\ion{O}{IV}] raises the possibility of some AGN contribution,
but the [\ion{O}{IV}]/[\ion{Ne}{II}] ratio restricts this to definitely less 
than 5 per cent.  
In addition to the fine structure recombination line ratios, the strengths
of the 6.2 $\umu$m PAH feature and the 9.7 $\umu$m silicate absorption 
feature ($S_{\rm SIL}$) place the Bird firmly in the starburst dominated 
ULIRG section of diagnostic diagrams in \citet{spoon07} and \citet{farrah07}.
  
Thus, if there is any AGN contribution from the Head (and this would perhaps be
the more unlikely location considering the other two nuclei are more 
massive) it is negligible compared to the total IR output of the Bird.
 
\begin{figure}
   \centering
   \includegraphics[width=84mm]{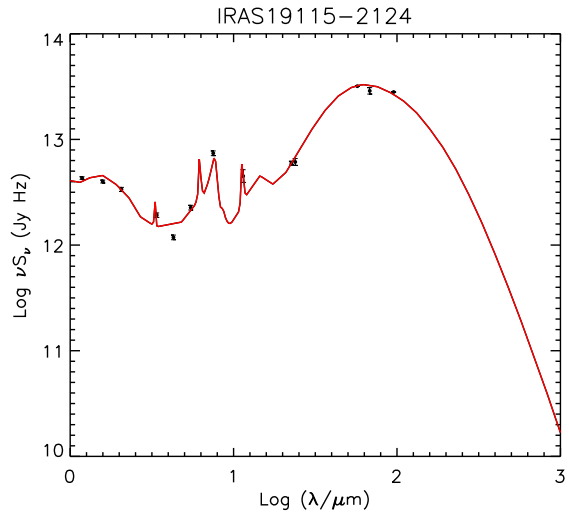}
   \caption{The 2MASS $JHK$, NACO $K$, {\em Spitzer} 3.6, 4.5, 5.6, 8, 24, 70 
    $\umu$m, and IRAS 12, 25, 60, 90 $\umu$m photometric 
    data overplotted with a model combining starburst and cirrus models. 
    See text for details.}
           \label{firsed}
    \end{figure}

We next estimate the star formation rate using the IR data available.
First, we have fitted models \citep{efs00, efs03} 
to the full IR SED of the Bird, from NIR 
NACO and 2MASS data, to the FIR IRS, MIPS, and {\em IRAS} data. 
Fig.~\ref{firsed} shows the data points and the SED model fit overplotted:
the best-fitting model, a pure starburst, 
assumes an underlying exponentially decaying SFR with a burst in the
last $7 \times 10^7$ years. The SFR at the peak of the burst was about 
400 ${\rm M}_{\odot} {\rm yr}^{-1}$ or 188 ${\rm M}_{\odot} {\rm yr}^{-1}$ 
when averaged over the last $7 \times 10^7$ years.

Second, we utilize a newly calibrated relation between star formation rate
and the PAH 6.2+11.2 $\umu$m luminosities $L_P$, based on correlations of these
PAH features with total IR luminosity \citep{farrah07}:
\[ SFR \ [{\rm M}_{\odot} {\rm yr}^{-1}] = 1.18 \times 10^{-41} L_P  \]
where $L_P$ is the  PAH 6.2+11.2 $\umu$m luminosity in ergs s$^{-1}$.  
From the IRS spectrum (Fig.~\ref{irs}) we calculate 
$L_P = 1.6 \times 10^{43} {\rm ergs \ s}^{-1}$,
i.e.\ we get a star formation rate of $SFR = 192 \ 
{\rm M_{\odot} {\rm yr}^{-1}}$, 
in very good agreement with the SED modelling.  For reference, we calculate 
a rate of $120 \ {\rm M}_{\odot} {\rm yr}^{-1}$  from the traditional 
\citet{kenni98} relation of ${\rm SFR} = 1.72 \times 10^{-10} L_{\rm IR}$ 
using the total IR luminosity of the Bird.

\subsection{Supernova rate}

Assuming core-collapse SN progenitors between 8 and 50 M$_\odot$ 
and a Salpeter 
initial mass function (IMF) with cut-offs at 0.1 and 125 M$_\odot$ 
\citep[see e.g.][]{seppo01}, we can estimate the core-collapse SN rate of 
IRAS 19115-2124. Adopting the average SFR of 190 M$_\odot$yr$^{-1}$ 
as obtained 
above, yields a core-collapse SN rate of about 1.3 yr$^{-1}$. 
Comparison of our NACO $K$-band images separated by 153 days did
not reveal any new sources within the galaxy nuclear regions 
(see Section~\ref{snsearch}) down to limiting magnitudes of $K = 21.3 - 21.5$. 
We compared these with template $K$-band light curves of ordinary 
(based on 11 type II and Ib/c SNe) 
and slowly-declining core-collapse SNe 
(based on the type IIL SN 1979C and the type IIn SN 1998S) 
from \citet{seppo01}. Assuming the Galactic extinction law \citep{riekeleb}
we find that any slow decliner, suffering from an 
extinction of up to A$_{V} \sim 30$, should have been detected if exploding 
between the epochs of the two images or less than a year before the first 
epoch image.  However, in the case of ordinary core-collapse SNe and the same
extinction, a detection would have been possible only near maximum light.
If the extinction was considerably lower, A$_{V}$ $\sim$ 5, we would have
detected also ordinary events anytime between the epochs of the two images 
or up to 5 months before the first epoch image.

Given the estimated core-collapse SN rate of 1.3 yr$^{-1}$ we estimate
a Poissonian probability of 16 per cent for a SN non-detection assuming that 
all the core-collapse SNe within the nuclear regions of the Bird were 
slowly-declining and suffered from extinctions of less than A$_{V}$ = 30. 
Recent observations of luminous radio SNe within nuclear regions of nearby
(U)LIRGs \citep{colina01, lonsdale06, alberdi06, parra07, pereztorres} 
show that 
these events have exploded 
within a dense circumstellar/interstellar environment.  
Therefore, a high rate of 
slowly-declining (in NIR) SNe might be expected in such environments.
However, if most of the SNe within LIRGs behave similarly to
SNe observed outside nuclear starburst regions, the fact that no SNe were 
detected is not surprising, since we were sensitive to ordinary core-collapse 
SNe with high extinctions only near the epoch of maximum light.

\subsection{Evolution of the Bird as a merger}
\label{evolution}

We next discuss possible evolutionary sequences of the 
Bird system aided by numerical merger simulations. Detailed dynamical
modelling of a given observed galaxy merger is very time consuming as a large
parameter-space of initial orientations and orbital configurations needs to be
probed \citep[e.g.][]{wahde,naab03,bar04,naab06}. Thus we here instead 
adopt an approach in which we model a typical 1:1 and a 3:1 galaxy merger in
order to study the general predictions from merger simulations. 

\begin{figure*}
\centering 
\includegraphics[width=16cm]{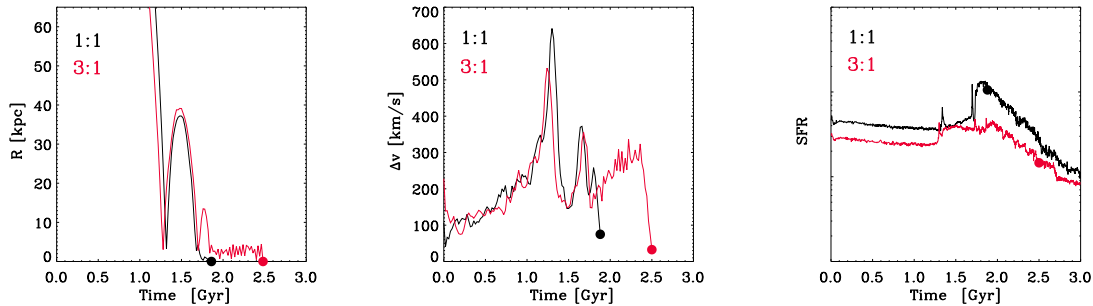}
\caption{Simulations of a 1:1 merger (black curves) and a 3:1 merger (red
curves). The left panel shows the nuclear separation, the middle panel the
relative velocity of the galaxy nuclei 
and the right panel the relative star formation rate of the total system, as a
function of time. The filled circle indicates the time of final coalescense of
the two galaxy nuclei.}
\label{ULIRG_sims}
\end{figure*}

The initial galaxy models are Milky Way-like galaxies with a disk mass of 
$m_{D}=5.5\times10^{10} {\rm M}_{\odot}$ and a 20 per cent initial 
gas fraction, the bulge mass
fraction is 
set to a third of the disk mass fraction and the galaxy is embedded 
in a dark matter halo with total virial mass of 
$m_{\rm{vir}}=1.34\times10^{12} {\rm M}_{\odot}$ (see
Johansson et al. 2007 in prep, \citet{springel05} for details). For
the 3:1 merger the corresponding masses for the minor component are a third 
smaller than the values given above.
The galaxies merge on a parabolic orbit with an initial separation of
$r_{\rm{sep}}=160 \rm{kpc} \ {\rm h}^{-1}$ and with a pericentric distance of  
$r_{\rm peri}=5.0 {\rm kpc} \ {\rm h}^{-1})$. 
The mergers are simulated with the paralell 
TreeSPH-code GADGET-2 \citep{springel} and the simulations follow
self-consistently the dark matter and gas dynamics, radiative gas cooling, star
formation, as well black hole growth and the associated
feedback processes \citep[Johansson et al. 2007 in prep.,][]{springel05}. 
Fig. \ref{ULIRG_sims} shows 
the resulting nuclear separations, relative velocities and corresponding total
relative star formation rates of the two mergers as a function of time.
 
Turning to our imaging data, 
the overall morphological appearance of the Bird, with multiple
identifiable nuclei within a few kpc, and very obvious tidal tails, 
classifies it as a close 'pre-merger', i.e.\ a system after first 
approach and contact (when the tidal tails typically are generated), 
but before a full merger, when nuclei are fully coalesced 
\citep[see e.g.][]{veil02}.  As for the individual nuclei, the NIR light in 
the inner regions of the 
Heart show a surprisingly intact structure with bulge and disk
components, and regular appearing barred 
spiral arms extending from the inner 300~pc out to some 2~kpc. The tidal
Wings beyond this region, however, break the regularity.  
Its rotation velocity vs.\ velocity dispersion ratio is in the
range  $v_{rot}/\sigma \sim 2 - 3$, i.e.\ a clear rotationally
supported disk system. The nucleus of the Body, in contrast, shows a much 
more disturbed morphology, with a surface brightness profile intermediate 
to that of disk and spheroidal systems. There is clear 
evidence of disturbed spiral arm and/or bar structure within the central 
1.3 kpc. It thus seems that the Body has thus far suffered more in the
on-going merger, and looks to be already morphing towards a 
de Vaucouleurs stellar light distribution.  The 
$v_{rot}/\sigma$ ratio is $\sim0.4$, which means that while it is a 
pressure supported system there is also significant rotational support
involved. 
The ratio $v_{rot}/\sigma \sim0.4$ appears to be a typical value 
in (U)LIRG samples \citep[e.g.][]{colina05, dasyra06a}, and the end 
result 
of the Bird merger will not be a slowly rotating massive elliptical, which are 
thought to be products of higher redshift galaxy formation 
\citep[e.g.][]{naab07}.

The relative (radial) velocity between the Heart and the Body was found to 
be very small, of the order of 50 km~s$^{-1}$, while the Head has a large 
positive velocity offset $\sim$400 km~s$^{-1}$ from the other two.  
Since also the strongest star formation is from the 
Head (see next section), a possible scenario is raised in which 
the Head is a foreground object {\em falling into}, or
just passing, the Body/Heart system. Comparing to Fig.~\ref{ULIRG_sims} we 
see that relative velocities in excess of 400 km~s$^{-1}$ are possible for
both 1:1 and 3:1 mergers; the Body/Heart:Head
system would more closely resemble a 3:1 merger. 
The peak in the relative velocity 
is associated with the first passage of the galaxies, which also produces a
clearly defined peak in the relative 
star formation rate at the same time of $t\sim 1.3
\ \rm{Gyr}$ (we did not attempt to model the level of SFR here, merely the
pattern as a function of time).
It is thus possible that the star formation of the whole system is dominated 
by the first triggering of star formation in the Head component, while the 
other nuclei are further into the merger and in between starburst phases. 
The locations of the bright NIR star clusters within the very extincted 
regions of the Head, at the edge closest to the other components, might be
an indication of just such infall of the component through the tidal ISM. 
One of the weak points of this scenario is 
the strikingly normal appearance of the Heart nucleus. However, given that the
Head is lower in mass by at least a factor of three than the Heart, it is
plausible that the Head is experiencing stronger tidal effects compared to the
Heart, leaving it more or less undisturbed.

Using the masses for individual components from Section~\ref{masses}, 
the minimum mass for the end result, derived from dynamics, would correspond 
to a $\sim 0.7 - 1.0 \ m_{\star}$ galaxy; total system mass from 
M/L relations is of the order  $\sim 2 \ m_{\star}$. 
The end result masses are quite high, though not exceptional.
Numerous studies \citep[e.g.][]{genzel01,colina05,dasyra06b,hinz06} 
indicate the end products of gas-rich spiral mergers, seen as LIRGs and 
ULIRGs, will be intermediate mass sub-$m_{\star}$ ellipticals.  The
IRAS~19115-2124 system is thus towards the high end of masses of local
IR bright systems, but nevertheless still well below the classification 
of a giant elliptical $>5 \times 10^{11} M_{\odot}$.

Finally, there are numerous bright concentrations within the Bird, and we have
not studied them in detail in this work. We note, however, that many of
them might be candidates for being, or becoming, tidal dwarf galaxies. 
As an example
the blue concentration within the Tail, 3.5 arcsec South of the Body,
is compact, distant, and massive (from NIR light) enough to satisfy the 
survival criteria for tidal dwarfs \citep{monreal07,wetzstein}. 

\subsubsection{Evolution of the Bird as a LIRG}

As far as the (U)LIRG phenomenon is concerned, the accummulated wisdom
indicates that the source of the huge IR luminosity comes from either AGN
or nuclear star formation, or both, and in the case of the Bird we have shown
that the star formation ovewhelmingly dominates.  For both  cases
the idea is that the central engine is fed by infalling gas from the
gas-rich progenitor disks, and that this infall is greatly facilitated by 
bars \citep[see e.g.][and references]{mihos,dimatteo}. 

Somewhat surprisingly, however, the majority of MIR output of the Bird system
is {\rm not} coming from either the major NIR bright galaxy nucleus (Body)
with its embedded bar-like structure, or from the barred spiral nucleus
(Heart).  The IRAC and MIPS imaging clearly indicate that the 
Head is the source of the bulk of MIR flux, with secondary
contribution from the Heart, while the contribution from the Body is 
negligible. 

The reason for non-conspicuous star formation in the major nuclei is not
the dearth of fuel for star formation; the CO observations clearly show that 
there is an ample reservoir of molecular hydrogen in all of the
major components (Fig.~\ref{mirafig}).  This would indicate that none of
the nuclei have yet experienced the second, or later, star formation peaks 
expected from merging models (see Fig~\ref{ULIRG_sims}).

There are recent reports of major off-nuclear and disk (rather
than centrally concentrated) star formation in interacting galaxies, 
\citep[][]{lipari00,wang04,jar06}, 
and even cases of non-enhanced or very moderate star formation 
connected with interactions \citep[e.g.][]{bergvall,cullen}. 
These studies highlight the fact that the triggering mechanism of 
star formation in interactions is still far from understood.  
All these cases are, however, in lower luminosity systems.
The Bird is thus quite unique in the sense that it is so close to being
classified as an ULIRG, and still is not dominated by a central 
starburst in the major components. Or, if the Head component were to be 
considered as a major nuclear location, the starburst is not associated with 
those gas-rich locations with bar structures.  

That strong star formation is present in the Head need not, of course, 
be surprising in itself; as discussed above, we may be witnessing the Head
interaction during its first passage by the two other nuclei that are further
along their merging sequence.  It is the timing which is more unique --
ULIRGs are typically observed in later stages of interactions, and
first passage starburst moments with associated large relative velocities are
rarely seen \citep[][]{murphy01}.

In summary, the existence of large gas reservoirs and connected bar structures 
in merging disk galaxies do not automatically imply a central star-burst
in that location.  On the other hand, general merger modelling 
does predict epochs of lower SFR in between more intense episodes.
In the case of the Bird it seems that the more massive nuclei are in such a  
phase, while the least massive component is caught virtually
in the act of first high-speed fly-by.
It would be very interesting to investigate further
what are the factors inhibiting the (U)LIRG phase in the Body nucleus, for 
example, or why exactly it has switched off.

\section{Summary of results and conclusions}

We have studied in detail the spectacular luminous infrared galaxy 
IRAS~19115-2124, which we have dubbed the Bird.  
Using near-diffraction limited adaptive optics imaging in the K-band
(VLT/NACO),
and matching space-based optical imaging (HST/ACS), combined with 
ground-based optical spectroscopy (SALT/RSS), we find the Bird to be
a pre-merger of three components. 

Two of the components have unambiguously identifiable 
galaxy nuclei.  One has the quite regular composition of a
barred spiral disk galaxy (Heart) while the other (Body) has
a more disturbed nuclear region, with hints of both spiral/arm
structure and a surface brightness profile intermediate to that 
of disks and ellipticals.  
The third component (Head) appears to be an irregular galaxy -- 
our kinematical and dynamical data, however, strongly suggest that it is not
merely a tidal dwarf borne out of the interaction of the other two, but
rather a pre-existing less massive galaxy intimately involved in the
interaction.  The AO NIR imaging is crucial in classifying sources and
nuclei due to the extremely clumpy nature of sources of 
this kind in the optical.  
In addition, we see strong tidal tails (the Wings) and 
numerous concentrated knots, some of which may be candidate TDGs, and
find a bright luminosity function for candidate SSCs.

The combination of high-resolution imaging and spectroscopy has resulted
in the refinement of the systemic velocity of this LIRG, 
$14576 \pm 9$ km~s$^{-1}$, and we have listed
the kinematics of all the components.  We have also investigated the line 
ratios, equivalent widths, and profiles of the important emission lines and
find them to show characteristics of an \ion{H}{II} type galaxy.  

We identify blueshifted \ion{Na}{I}~D absorption doublets mainly
from the Heart, Head, and Body components.  We interpret these to be due to 
gas outflows from the Bird, with velocities typically in the range 
$100-300$ km~s$^{-1}$, but possibly some flows ranging up to 
600 km~s$^{-1}$.  Consistently, we also identify 
elevated [\ion{N}{II}]/H$\alpha$ levels in blueshifted wings of 
emission lines from the nuclear components, suggesting shock-heating
associated with outflows.

By utilizing {\em Spitzer} IRAC and MIPS imaging, we are able to pinpoint where
the MIR flux and the non-stellar emission is originating
from.  The Heart and especially the 
Head components are the regions producing most
of the MIR flux.  
There is a hotter dust component affecting the $4-8\umu$m bands
at the location of the Head: {\em Spitzer} IRS spectroscopy is used to 
confirm the classification suggested by our optical spectroscopy,
that this location, and the Bird system as a whole, 
is a star formation dominated system, with any AGN contribution below 
5 per cent.  The star formation is estimated to be 
$\sim 190 \ {\rm M_{\odot} yr^{-1}}$,
corresponding to a core-collapse supernova rate of $\sim$1.3 yr$^{-1}$. 
We detected no supernovae brighter than $K\sim21$ in 
NIR images separated by 5 months.

We estimate averaged extinction over the system to be of the order 
A$_V \approx 3-5$, while the extinction derived from MIR diagnostics of the
deeply embedded nuclear system results in A$_V \approx 30$. Gas masses
inferred are up to a level of $M \sim 2 \times 10^9 {\rm M}_{\odot}$.

We have determined the component 
masses of the Bird system using rotational as well as
central velocity dispersion information. We note that without the 
high-resolution NIR imaging it would have been very difficult to differentiate
between velocity components related to nuclear bulges from those
originating e.g.\ from
\ion{H}{II} regions in tidal regions or outflow components. Using conservative
deblending of components we arrive at masses of 0.27~$m_{\star}$, 
0.36~$m_{\star}$ and 0.08~$m_{\star}$ for the Body, Heart, and Head,
respectively. Alternatively, using K-band light as a tracer, we find masses
of 0.40~$m_{\star}$, 0.25~$m_{\star}$ and 0.16~$m_{\star}$ 
for the inner 1~kpc regions of the same three regions, and 2.1~$m_{\star}$
for the total system.

In summary, most optical, NIR, MIR, and FIR emission characteristics, as well
as kinematical and dynamical attributes place IRAS~19115-2124 among typical 
ULIRG samples, even though it is not technically quite a ULIRG with an IR 
luminosity of $L_{IR} = 10^{11.9} {\rm L}_{\odot}$.  
The main differences are that 
it appears to be a three-component merger, and that its total dynamical mass 
is of the order of $\sim 1 m_{\star}$, 
or more, i.e.\ at the upper end of typical 
ULIRG merger remnants.  Finally, we find 
the likely origin of the ULIRG phase engine to be the Head, the least 
massive of the three merger components, and the only one without signs of
spiral arms and bars, which often are regarded as tell-tale signs of
inflowing gas to central starburst regions. A possible sequence of events
is such that the Heart and Body met first; the Body especially is already
in the process of morphing into an elliptical like structure.  The Head
is observed during its first high-speed approach to the system, making it
the strongest starburst at this time -- the general characteristics of this
scenario are consistent with numerical merger simulations.

\section*{Acknowledgments}

SM acknowledges financial support from funds from the Participating
Organisations of EURYI and the EC Sixth Framework Programme and from
the Academy of Finland (project: 8120503).  
  Some of the observations reported in this paper were obtained with
  the Southern African Large Telescope (SALT), a consortium consisting 
  of the National Research Foundation of South Africa,
  Nicholas Copernicus Astronomical Center of the Polish Academy of Sciences, 
  Hobby Eberly Telescope Founding Institutions, Rutgers University, 
  Georg-August-Universit\"at G\"ottingen, University of Wisconsin - Madison, 
  Carnegie Mellon University, University of Canterbury, United Kingdom SALT
  Consortium, University of North Carolina - Chapel Hill, Dartmouth College, 
  American Museum of Natural History and the Inter-University Centre for 
  Astronomoy and Astrophysics, India.
This work is based in part
      on observations with the NASA/ESA Hubble Space Telescope, 
      obtained from the data archive at the Space Telescope Institute. STScI 
      is operated by the association of Universities for Research in Astronomy,
      Inc. under the NASA contract  NAS 5-26555, and based in part on 
archival data 
      obtained with the Spitzer Space Telescope, which is operated by the 
      Jet Propulsion Laboratory, California Institute of Technology under 
      a contract with NASA. 
This research has made use of the NASA/ IPAC 
      Infrared Science Archive, which is operated by the Jet Propulsion 
      Laboratory, California Institute of Technology, under contract with 
      the National Aeronautics and Space Administration, and made 
   use of the NASA/IPAC Extragalactic Database (NED) which is operated by 
   the Jet Propulsion Laboratory, California Institute of Technology, under 
   contract with the National Aeronautics and Space Administration.

\appendix

\bsp

\label{lastpage}

\end{document}